\documentclass[a4paper,12pt]{article}
\usepackage[cmex10]{amsmath}
\usepackage{latexsym,amssymb}
\usepackage[mathscr]{eucal}
\usepackage{amsxtra,amscd,amsthm,upref,layout,color}
\usepackage{calc}
\usepackage[final]{graphicx}
\def\refup#1{{$^{#1}$}}
\def\p{^{^{\prime}}}\def\pp{^{^{\prime\prime}}}
\def\ppp{^{^{\prime\prime\prime}}}
\def\ppp{^{^{\prime\prime\prime}}}

\def\br{{\bf r}}

\def\begeq{\begin{equation}}
\def\endeq{\end{equation}}
\def\begdis{\begin{displaymath}}
\def\enddis{\end{displaymath}}
\def\cA{{\cal A}}\def\cC{{\cal C}}\def\cG{{\cal G}}\def\cGh{\hat{\cG}}
  \def\cI{{\cal I}} 
\def\cQ{{\cal Q}}    \def\cP{{\cal P}}
\def\cK{{\cal K}}  \def\cR{{\cal R}}\def\cS{{\cal S}}
\def\cV{{\cal V}}
\def\cQS1{{\cQ_{\cS_1}}}

\def\prd{{\rm P}_{rd}}
\def\ie{{\em i.e.}}
\def\etal{{\em et al.}}
\def\cA{{\cal A}}\def\cC{{\cal C}}
\def\br{{\bf r}}
\def\br{{\bf r}}\def\hw{{\hat \omega}}
\def\gr{\gamma(r)}
    
\def\hw{{\hat \omega }}

\def\mp{M_{_{\rm P}}}
\def\href#1{\relax}
\begin{document}
\title{Generalization of  the Fedorova-Schmidt method for determining  particle size distributions}
\author{
{{Salvino Ciccariello}}
\\[6mm]
\begin{minipage}[t]{3mm}
           \
\end{minipage}
\begin{minipage}[t]{0.9\textwidth}
  \begin{flushleft}
  \setlength{\baselineskip}{12pt}
  {\slshape{\footnotesize{
  Universit\`{a} di Padova, Dipartimento di Fisica {\em G. Galilei}
  }}}\\{\slshape{\footnotesize{
  Via Marzolo 8, I-35131 Padova, Italy.
  }}}\\{\slshape{\footnotesize{
  E-mail {\upshape{\texttt{salvino.ciccariello@unipd.it}}}; Phone +39\,0473\,690808;\\ 
Fax +39~049~8277102
  }}}
  \end{flushleft}
\end{minipage}\\[10mm]
}

\date{ July 21, 2014}

\maketitle                        
\begin{abstract}
\setlength{\baselineskip}{18truept}
\noindent  
One reports the integral transform that determines the particle size distribution of a given sample 
from the small-angle scattering  intensity under the assumption that the particle correlation
function is a polynomial of degree $M$. The Fedorova-Schmidt solution [{\em J. Appl. Cryst.} 
{\bf 11}, 405, (1978)] corresponds to the case $M=3$. The procedure for obtaining a 
polynomial approximation to a particle correlation function is discussed and applied 
to the cases of polidisperse particles of  tetrahedral or octahedral or  cubical shape. 
 \\    
\\  
\\ 
Synopsis: It is reported the integral transform that determines the particle size distribution 
from the small angle scattering intensity under the assumption that the particle correlation 
function is a polynomial. 
\end{abstract}
\newpage      
\setlength{\baselineskip}{12pt}
                 \section{Introduction}
Colloidal suspensions, demixing alloys, porous glasses, carbons are
classical examples of materials that have a particulate structure
on a length scale of 1-$10^3$nm. For these samples
scattering data can successfully be used to determine the
particle size distribution in the favorable cases where:  a) the constituting
particles and the surrounding medium are fairly homogeneous on a spatial
resolution of 1nm, b) the sample particles have the same geometrical shape
and are isotropically distributed, and c) interference effects among different
particles are negligible. The application of this procedure to small-angle
scattering (SAS) intensities traces back to the contributions of
Roess(1946) and Roess \& Shull (1947). In particular,  Roess \& Shull 
devised an early scheme whereby the particle size distribution could be
found by comparing the observed SAS intensity with plots of
theoretical intensities numerically evaluated by considering
reasonable particle shape and appropriate size-distribution
functions. Besides,  Roess (1946) also showed that  the size
distribution can be expressed as an integral transform of the
observed intensity in the case of spherical shape. This interesting result
was improved and successfully applied by Letcher \& Schmidt(1966) to
get the particle size distributions of three Ludox samples by the analysis
of  their small-angle x-ray (SAXS) intensities. Some years later
Fedorova \& Schmidt (1978) generalized the previous result  showing that
the size distribution can be expressed as an integral
transform of the SAS intensity for all the particle shapes
characterized by an isotropic form factor of the following form
\begeq\label{1.1}
{{[v\,\ell^{\alpha}J_{\nu}(q\,\ell)]^2}\over{(q\,\ell)^{\beta}}}
\endeq
where $q=(4\pi/\lambda)\sin(\theta/2)$ is the scattering vector (
$\lambda$ denoting the  particle beam  wave-length and $\theta$ the
scattering angle),
$\ell$ denotes the maximal chord of the particle,  $v\,\ell^{\alpha}$  its volume, 
$\nu,\ \alpha$ and $\beta$ depend on the particle  shape and dimensionality, 
and  $J_{\nu}(\cdot)$ is the Bessel function of the 1st kind. The shapes
allowed by this more general formulation include hollow spheres, disks of
negligible thickness and  uniform or hollow cylinders of infinite length.
Besides, in all these cases, Fedorova and Schmidt worked out  the integral
transform that yields  the size distribution from the SAS
intensity collected with the pin hole or the infinite slit geometry. \\ 
The main aim of this  paper is to get a generalization of this result.\\ 
In fact, it will be shown that the size distribution can be written as an
integral transform of the collected intensity in all the cases where the particle 
correlation function is an $M$ degree polynomial $P_M(r)$ continuous,  
together with its  first $m(<M)$ derivatives, throughout $[0,\,\infty]$. 
Consider a particle of maximal chord $\ell$.  The correlation function (CF) 
of this particle identically vanishes once $r$ exceeds $\ell$. Hence, the 
above assumption of continuity is equivalent to state that the CF and its 
first $m$ derivatives vanish at $r=\ell$.  \\
The plan of the paper is as follows. For completeness, section 2 reports 
the expression of the correlation function (CF) in terms of the so-called stick 
probability functions (Debye \etal\, 1957; Goodisman \& Brumberger, 1971) while  
\S\,3 expresses the observed scattering intensity in terms of the particle CF, the 
particle size distribution and a further contribution related to the overlapping of 
different particles. The Fourier transform of this contribution, on the average,  
negatively contributes to the observed scattering intensity and vanishes in the infinite dilution limit. Hence, as firstly pointed out by Guinier and  Fournet (1955), the 
relation exploited by polidisperse analysis is physically consistent only for 
highly diluted samples.  Sections 4 and section 5 respectively report the 
generalization of the Fedorova-Schmidt result in direct and in reciprocal space.  
Section 6  first discusses how  to construct a polynomial approximation of a  particle CF, and then, in \S 6.1, 6.2.1 and 6.2.2, analyzes the cases of the sphere, the cube/octahedron and  the 
tetrahedron.
  The final conclusions are reported in \S\,7.
     \section{Stick probability functions and correlation function}
The scattering intensity $I(q)$ is the square modulus of the
Fourier  transform (FT) of $n(\br)$, the scattering density of the sample.
In the SAS domain,  $n(\br)$ is fairly approximated by a two value function that
can therefore be written as
\begeq\label{2.1}
n(\br)=n_1\rho_1(\br)+n_2\rho_2(\br)
\endeq
where $n_1$ ($n_2$) denotes the scattering density of phase 1 (2) and $\rho_1(\br)$ 
defines the full geometry of phase 1 since it is equal to 1 or 0 depending on whether the tip of $\br$ falls inside or outside phase 1.  (The definition of $\rho_2(\br)$ is quite similar.)  The spatial region occupied by the $i$th phase will
be denoted as $\cV_i$ and its volume by $V_i$. The volume of the sample is $V=V_1+V_2$
and the ratio $\varphi_i\equiv V_i/V$ is the volume fraction of the $i$th phase. According to
Debye \etal (1957) the {\em stick probability function} relevant
to the $i$th and $j$th phase is defined as
\begin{equation}\label{2.2}
P_{i,j}(r)={1\over{4\pi V}}\int d\hw \int {\bf \rho_{i}}(\br_1){\bf \rho_{j}}
(\br_1+r\hw)dv_1,\quad i,\,j=1,\,2.
\end{equation}
Here $\hw$ is a unit vector that spans all possible directions. Function $P_{i,j}(r)$ is the 
angular average of  the volume fraction of the overlapping region between phases $i$ and $j$, once the latter has been translated by $-r\hw$. It also represents the probability that 
a stick of length $r$, after having been randomly tossed a very large number of times, falls 
with one end inside phase $i$ and the other end within phase $j$. The
$P_{i,j}(r)$s have the following properties [ Goodisman \& Brumberger (1971), Ciccariello \etal (1981), Ciccariello  (1984)]
\begeq\label{2.3}
P_{i,j}(r)=P_{j,i}(r),\quad P_{i,1}(r)+P_{i,2}(r)=\varphi_i,
\endeq
\begeq\label{2.4a}
P_{i,j}(0)=\varphi_i \delta_{i,j},\quad P_{i,j}(\infty)=\varphi_i\varphi_j,\quad
P_{i,j}\p(0)=(-1)^{i+j}S/4V,
\endeq
\begeq\label{2.4b}
P_{i,j}\pp(0)=(-1)^{i+j}\sum_{j}\frac{L_j}{{3\pi V}}\left(1+\left(\pi-\beta_j\right)\cot\,
\beta_j\right),\endeq
\begeq\label{2.4c}
P_{i,j}\ppp(0)=(-1)^{i+j}\frac{1}{16V}\int_S\left(3\,H^2(\br)-K_G(\br)\right)dS +\frac{{{\sf S}_{i,j}}}{4V}.
\endeq
Here  the prime  denotes the derivative, $S$ the area of the interphase surface,
 $L_j$ the length of the $j$th edge present on the interface,  $\beta_j$ the associated dihedral angle.  Further, in (\ref{2.4c}), $H(\br)$ and $K_G(\br)$ respectively denote the mean and the Gaussian curvature of the interface at the point $\br$ and ${\sf S}_{i,j}$ a further contribution that is only present when the interphase surface presents edges meeting at some vertices. We defer to  Kirste and Porod (1962) for the derivation of 
the integral contribution and to  Ciccariello \& Sobry (1995) for the explicit expression
of ${\sf S}_{i,j}$. \\
For a statistically isotropic sample one finds that [Debye \etal (1957), Goodisman \&
Brumberger (1971), Ciccariello \etal (1981)] the scattering intensity $I(q)$ is
\begin{equation}\label{2.5}
I(q)={{4\pi}\over{q}}\langle\eta^2\rangle V\int_0^{\infty}r\sin(qr)\Gamma(r)dr
\end{equation}
with
\begin{equation}\label{2.6}
\Gamma(r)\equiv 1-(n_1-n_2)^2 P_{1,2}(r)/\langle\eta^2\rangle,
\end{equation}
where $\langle\eta^2\rangle$ is the mean square scattering density fluctuation, \ie
\begeq\label{2.7}
\langle\eta^2\rangle\equiv\sum_{i=1}^2 (n_i-{\bar n})^2\varphi_i=(n_1-n_2)^2\varphi_1\varphi_2
\endeq
with ${\bar n}\equiv (n_1\varphi_1+n_2\varphi_2)$.
Function $\Gamma(r)$ is commonly referred to as the correlation function of the
sample. By equation. (\ref{2.3}b) it takes the form
\begin{equation}\label{2.8}
\Gamma(r)={{P_{1,1}(r)-\phi_1^2}\over{\phi_1\phi_2}},
\end{equation}
more useful in our later analysis. From (\ref{2.4a}) and (\ref{2.8}) immediately
follows that
\begeq\label{2.9}
\Gamma(0)=1,\quad \Gamma(\infty)=0,\quad\
\Gamma\p(0)=-S/(4V\varphi_1\varphi_2).
\endeq
Similarly, the expressions of $\Gamma\pp(0)$ and $\Gamma\ppp(0)$ are also
known by Equation.s (\ref{2.4b}) and (\ref{2.4c}).\\
Multiplying $I(q)$ by $4\pi\,q^2$,  integrating over all the positive $q$s and 
using Equation. (\ref{2.5}) and (\ref{2.9}a), one finds the so-called Porod invariant relation
\begeq\label{2.10}
\cQ_P\equiv \int_0^{\infty}q^2I(q)\,dq=2\,\pi^2\,V\langle\eta^2\rangle .
\endeq

              \section{The scattering intensity in the polidisperse approximation}
Consider now the case of a particulate sample where the particles have the same
shape and different sizes. The distribution of the particles in the space is
assumed to be statistically isotropic. The phase formed by the particles will be
named phase 1. Hence $\rho_1(\br)$ becomes equal to
\begeq\label{3.1}
\rho_{1}(\br)=\sum_{J}\rho_J(\br),
\endeq
where the sum runs over all the particles, labelled by index $J$ and $\rho_J(\br)$  denotes now the characteristic function of the $J$th particle. 
The substitution of (\ref{3.1}) into  the $P_{1,1}(r)$ definition [see equation (\ref{2.2})] yields  
\begeq\label{3.2} 
P_{{1,1}}(r)=   
\sum_{J}{{V_J\gamma_J(r)}\over{V}}+\cR(r),
\endeq
with 
\begeq\label{3.2a}
\cR(r)\equiv 
\sum_{J\ne L}{1\over{4\pi V}}\int d\hw\int
\rho_J(\br_1)\rho_L(\br_1+r\hw)dv_1,
\endeq 
\begeq\label{3.3}
\gamma_J(r)\equiv {1\over{4\pi V_j}}
\int d\hw\int\rho_J(\br_1)\rho_J(\br_1+r\hw)dv_1,
\endeq
and $V_J$ denoting the volume of the $J$th particle.  
The sum on the right hand side (rhs) of (\ref{3.2}) and function $\cR(r)$ respectively are the intraparticle and the interparticle contributions to $P_{{1,1}}(r)$. The last contribution is poorly known. One knows that it and its first derivative vanish as $r\to 0$ (because  the overlapping between two different particles is possible only if $r$ is greater than the lowest of the particle maximal chord values)  and  that it approaches $\varphi_1^2$  as $r$ becomes very large  (because  the probability that the stick of length $r$ has 
its ends within phase 1 is equal to $\varphi_1^2$). \\ 
We elaborate now the intraparticle contribution. To this aim 
one observes that function $\gamma_J(r)$ 
 is independent on the actual position and orientation of the particle and only
depends on the shape (fixed by assumption) and the size of the particle. It is the (isotropic) correlation function of the $J$th particle. It is equal to zero as $r$ exceeds $\ell_J$, the largest chord of the particle, and is equal to one at $r=0$. Comparing $\gamma_J(r)$ with $\gamma_L(r)$, it results that $\gamma_J(r)=\gamma_L(\ell_L\,r/\ell_J)$. 
Then, denoting by $\gamma(r)$ the CF of the {\em unit} particle, namely the particle having   its largest chord equal to the unit length, one has  
\begeq\label{3.4}
\gamma_J(r)=\gamma(r/\ell_J).
\endeq
 The volume of the $J$th particle is  $V_J=v\,\ell_J^3$ and the first sum on the rhs of (\ref{3.2}) becomes
\begeq\label{3.5}
\sum_{J}\frac{V_J\gamma_J(r)}{V}=v\sum_j\frac {N_j\,\ell_j^3}{V}\gamma(r/\ell_j)=
\frac{v\, N_t}{V}\sum_j\frac {{N_j}}{N_t}{\ell_j}^3\,\gamma(r/\ell_j),
\endeq
where the sum runs now over the particles'  different sizes labelled by $j$. Moreover,
we have denoted  by $N_j$ the number of  particles with the $j$th size, and  by $N_t$
the total number of the  particles present in the sample.  As $V$ becomes 
infinitely large,  $N_t/V$ becomes equal to $n$, the particle number mean density
of the sample and, what is more important, since $j$ runs over a very large
number, $N_j/N_t$  can be approximated  by the  infinitesimal quantity
$p(\ell)d\ell$ where $p(\ell)$ represents the  probability density of finding a particle
of size $\ell$ within the interval $[\ell,\,  \ell+d\ell$]. Then,  the first sum on the
rhs of (\ref{3.2}) converts into  an integral, \ie\
\begeq\label{3.6}
  \sum_{J}\frac{V_J\gamma_J(r)}{V}\ \to \ v\,n \int_0^{\infty}\ell^3
\,p(\ell)\,\gamma(r/\ell)\,d\ell = v\,n \int_r^{\infty}\ell^3
\,p(\ell)\,\gamma(r/\ell)\,d\ell.
\endeq
From equations  (\ref{2.8})) and (\ref{3.2}) one finds that the sample CF is the sum of two  terms 
\begeq\label{3.7}
\Gamma(r)=\Gamma_p (r) + \Gamma_i(r),
\endeq 
with 
\begeq\label{3.7a}
\Gamma_p (r)\equiv \frac{\ v\,n}{\varphi_1\varphi_2} \int_0^{\infty}\ell^3\,p(\ell)\,\gamma(r/\ell)\,d\ell
\endeq 
and 
\begeq\label{3.7b}
\Gamma_i(r)\equiv \frac{\cR(r)-\varphi_1^2}{\varphi_1\varphi_2}.
\endeq 
The CF's definition requires that   $\Gamma(0)=1$.  Setting 
$r=0$ into (\ref{3.7a}) and observing that the particle mean volume ${\bar v}_p$ is given by \begeq\label{3.12a}
{\overline v_p}=v \int_0^{\infty}\ell^3p(\ell)d\ell 
\endeq
so that $\varphi_1=n\,{\overline v_p}$, one finds that      
\begeq\label{3.7x}
\Gamma_p(0)=\frac{v\,n}{\varphi_1\,\varphi_2} \int_0^{\infty}\ell^3\,p(\ell)\,d\,\ell
=\frac{n {\bar v_p}}{\varphi_1\,\varphi_2} =\frac{1}{\varphi_2}.
\endeq 
 Since $\cR(0)=0$, one also finds that 
\begeq\label{3.7c}
\Gamma_i(0)=-\varphi_1/\varphi_2.
\endeq 
From (\ref{3.7x}) and (\ref{3.7c}) one concludes that neither $\Gamma_p(r)$ nor 
$\Gamma_i(r)$ are CFs and that the only $\Gamma(r)$ shares this property because 
$\Gamma(0)=1$. 
By Fourier transforming  relation (\ref{3.7}), multiplied by $\langle\eta^2\rangle V$, and putting  
\begeq\label{3.9a}
\cC\equiv (n_1-n_2)^2 \,v \,n \,V,
\endeq
one finds that the scattering intensity, given by  (\ref{2.5}), separates into two contributions, \ie
\begeq\label{3.8}
I(q)=I_p(q)+I_i(q)
\endeq
with    
\begeq\label{3.9}
I_p(q)\equiv \frac{4\pi \cC} {q} \int_0^{\infty} \ell^3 p(\ell) d\ell
\int_0^{\infty}r\,\sin(qr)\gamma(r/\ell)dr,
\endeq             
and
\begeq\label{3.10}
I_i(q)\equiv\frac{4\,\pi\,\cC} {q} \int_0^{\infty} r\,\sin(q\,r)\left(\frac{R(r)-\varphi_1^2}{\varphi_1\,\varphi_2}\right)dr.
\endeq
$I_p(q)$ and $I_i(q)$ respectively represent the intraparticle and the interparticle contribution  to the scattering intensity.  It is noted that Guinier \& Fournet (1955) first  
discussed the importance of the contribution $I_i(q)$, which was mainly analyzed in 
the  molecular  approximation.  Apply now the basic relation 
\begeq\label{3.10a}
\frac{1}{2\pi^2}\int_0^{\infty}q^2\,d\,q\left(\frac{4\pi}{q}\int_0^{\infty}r\sin(q\,r)f(r)dr\right)=f(0), 
\endeq 
 to $I_p(q)$ and $I_i(q)$. One respectively finds 
\begin{eqnarray}\label{3.10b}
&& \cQ_{p,P}\equiv\int_0^{\infty}q^2\, I_p(q)\,d\,q=2\,\pi^2\langle\eta^2\rangle V\Gamma_p(0)=\\
&&\quad \quad\quad\quad
\frac{2\,\pi^2\,\langle\eta^2\rangle V}{\varphi_2}=
2\,\pi^2\,(n_1-n_2)^2\,N_t{\overline v_p}\nonumber
\end{eqnarray} 
and 
\begin{eqnarray}\label{3.10c}
&&\cQ_{i,P}\equiv\int_0^{\infty}q^2\, I_i(q)\,d\,q=2\,\pi^2\langle\eta^2\rangle V\Gamma_i(0)=\\
&& \quad \quad\quad\quad
-\frac{2\,\pi^2\,\langle\eta^2\rangle V\,\varphi_1}{\varphi_2} =- 2\,\pi^2\,(n_1-n_2)^2\,N_t\, n\,{\overline v_p}^2<0.   \nonumber
\end{eqnarray}     
The negativeness of $\cQ_{i,P}$ implies that $ I_i(q)$ is negative in some $q$-ranges 
so that $ I_i(q)$ is not a scattering intensity but only a FT. Owing to the 
fact that $\cR\p(r)$ also tends to zero  as $r\to 0$, it follows that 
$I_i(q)$ decreases faster than $q^{-4}$ as $q\to\infty$. One concludes that the 
asymptotic behaviour of $I(q)$ is equal to that of $I_p(q)$. The last one immediately follows from (\ref{3.9}) and from the fact that
$\gamma\p(0) = -s/4v$, where $s$ denotes the surface area of the unit particle.  One finds that Porod's law (Porod, 1951) takes the form 
\begeq\label{3.15}
I(q)\approx I_p(q)\approx \frac{2\pi (n_1-n_2)^2 \,s \,N_t } {q^4} \int_0^{\infty} \ell^2 p(\ell) d\ell=
\frac{2\pi (n_1-n_2)^2\,N_t \,{{\bar s}_p} } {q^4}=\frac{\prd}{q^4},
\endeq
with 
\begeq\label{3.16}
{\bar s}_p\equiv s\int_0^{\infty} \ell^2 p(\ell) d\ell
\endeq
denoting the particle mean surface area, ${\bar S}\equiv N_t\,{\overline {s}}_p$ and $\prd\equiv 2\pi (n_1-n_2)^2 \,{\bar S} $.  \\ 
 Moreover, the only sum of (\ref{3.10b}) and (\ref{3.10c})  yields the correct  Porod invariant  [\ie\, (\ref{2.10})]. \\ 
The polidisperse approximation consists in setting $I_p(q)\approx I(q)$ and 
neglecting $I_i(q)$ (Guinier \& Fournet, 1955; Feigin \& Svergun, ; Gille, 2013). The approximation is certainly correct at large $q$s for the behaviour of $I_i(q)$. 
The last  contribution  is only appreciable at small $q$s where the approximation   $I_p(q)\approx I(q)$ consequently fails unless the sample is very dilute, because  $I_i(q)$ 
vanishes as  $\varphi_1\to 0$\footnote{A heuristic way for overcoming the high dilution  assumption consists in assuming that $\cR(r)$ has a known parameterized form, for instance $\cR(r)=\varphi_1^2(1-(1+\mu\,r)e^{-\mu\,r}\cos(\nu\,r))$, which vanishes together 
with its derivative as $r\to 0$ and tends to $\varphi_1^2$ as $r\to\infty$. Then  $\Gamma_i(r)$ is obtained by (\ref{3.7b}), $I_i(q)$ by (\ref{3.10}) and $I_p(q)$ is 
known (in terms of $\mu$ and $\nu)$ by the relation $I_p(q)=I(q)-I_i(q)$. Finally the four  
relations (\ref{2.10}), (\ref{3.10b}),  (\ref{3.10c}) and (\ref{3.15}) in principle determine, $(n_1-n_2)^2$, $N_t$, $\mu$ and $\nu$. }. Only in this case 
 $I_p(q)$ fairly fulfills the Porod invariant relation because the rhs of (\ref{3.10b}) closely approaches   $2\pi^2\langle\eta^2\rangle V$.\\ 
Finally,  it is observed that the knowledge of 
$\cQ_{p,P}$,  $\prd$ and $p(r)$ only determines $(n_1-n_2)^2\,N_t$ since one has    $(n_1-n_2)^2\,N_t=\frac{\cQ_{p,P}}{2\pi^2{\bar v}_p}=\frac{\prd}{2\pi^2{\bar s}_p}$. 
From the last equality one obtains that $\frac{{\bar s}_p}{{\bar v}_p}=\frac{\prd}{\cQ_{p,P}}$ which tests the accuracy of $p(r)$.

                           \section {Generalization of the Fedorova-Schmidt method}
Hereafter it will be assumed that the approximation $I_p(q)\approx I(q)$ holds true and 
the problem of obtaining an integral transform that directly determines $p(\ell)$ from the scattering intensity $ I(q)$ will now be tackled.  
In principle, if $I(q)$ is known throughout the full $q$-range, function $G(r)$,  
defined as 
\begeq\label{4.1a}
G(r)\equiv \frac{1}{2\pi^2\, \cC}
\int_0^{\infty}q\sin(qr)I(q)dq,
\endeq 
also is fully known. On the other hand, by equation (\ref{3.9}) one finds that 
\begeq\label{4.1}
\frac{1}{2\pi^2\, \cC}
\int_0^{\infty}q\sin(qr)I(q)dq = \int_0^{\infty} \ell^3 p(\ell) \gamma(r/\ell)d\ell.
\endeq 
 Combining (\ref{4.1a}) and (\ref{4.1}) and putting, for notational simplicity,   
\begeq\label{4.1c}
\cP(\ell)\equiv \ell^3p(\ell), 
\endeq 
one gets the integral  equation 
\begeq\label{4.1d}
\int_0^{\infty} \cP(\ell) \gamma(r/\ell)d\ell=G(r),
\endeq
which, once solved,  determines $\cP(\ell)$, and hence $p(\ell)$ {\em via} (\ref{4.1c}), in 
terms of $G(r)$,  and hence of $I(q)$ {\em via} (\ref{4.1a}).  Using the property 
that $\gamma(r)$ identically vanishes if $r>1$,  equation (\ref{4.1d}) converts into
\begeq\label{4.2}
G(r)=\int_r^{\infty} \cP(\ell) \gamma(r/\ell)d\ell.
\endeq                                     
Our task now is that of showing  that  the (integral) transform that expresses 
$\cP(\ell)$ in terms  $\gamma(r)$ and $G(r)$ can be explicitly 
written down  if one assumes that:
\begin{itemize}
\item {\em  i)} $\gamma(r)$ {\em  is an $M$th degree polynomial},
\ie\
\begeq\label{4.3}
\gamma(r)=\sum_{i=0}^M a_i\, r^i, \quad M \ge 3, 
\endeq
 with $a_0=1$ and $a_1=-s/4v$,
\item{{\em ii)}}  $\gamma(r)$ {\em and its  first $m$ derivatives (with $1\le m<M$) 
vanish at $r=1$}, \ie\ after putting
\begeq\label{4.3a}
g_m\equiv\gamma^{(m)}(1)\quad{\rm for}\quad m=0,1,\ldots,M,
\endeq
one assumes that
\begeq\label{4.4}
g_0=g_1=\ldots=g_m=0,
\endeq
\item{{\em iii)}} $\cP(\ell)$ {\em exponentially decreases as 
$\ell\to\infty$ and, as $\ell\to 0$, it goes to zero  sufficiently fast for the
later considered  integrals to exist}. 
\end{itemize}
Owing to assumption {\em ii)},  from (\ref{4.2})  follows that 
\begeq\label{4.5}
G^{(k)}(r) =  \int_r^{\infty} \ell^{-k} \cP(\ell) \gamma^{(k)}(r/\ell)d\ell\quad
{\rm if}\quad 0\,\le\, k\,\le\, m+1.
\endeq   
For the $(m+2)$th derivative one can no longer use (\ref{4.4}) and one finds that 
\begeq\label{4.6}
G^{(m+2)}(r) = -g_{m+1}\, \cP(r) /r^{m+1}+ \int_r^{\infty}
\ell^{-m-2}\, \cP(\ell)\,\gamma^{(m+2)}(r/\ell)\,d\ell.
\endeq
Up to the $M$th order, the successive derivatives have a similar structure. 
The $(M+1)$th derivative however will involve no integral contribution because 
the integrand vanishes owing to assumption {\em i)}. In this way one finds that  
$\cP(r)$ must obey the following linear inhomogeneous differential equation
\begeq\label{4.7}
\sum_{k=0}^{(M-m-1)}g_{M-k}\frac{d^{k}\ }{dr^{k}}\,\left(\cP(r)/r^{M-k}\right)
=\,- G^{(M+1)}(r).
\endeq
If $m=(M-1)$ the sum on the left hand side reduces to a single term so that
$\cP(r)$ is simply determined by
\begeq\label{4.7a}
\cP(r)=- r^{m+1}G^{(M+1)}(r)/g_{m+1}.
\endeq
If $m<(M-1)$, Equation. (\ref{4.7}) is a genuine  differential equation of
order $\mp\equiv(M-m-1)$ yielding $\cP(r)$ as one of its solutions.  \\
To get $\cP(r)$  one must first determine
the general integral of the associated homogeneous differential equation
\begeq\label{4.8}
\sum_{k=0}^{\mp}g_{M-k}\frac{d^{k}\ }{dr^{k}}
\left(\cP(r)/r^{M-k}\right) =0,
\endeq
then find a particular integral of (\ref{4.7}) and, finally, impose the appropriate 
boundary conditions.  \\
Looking for a solution of (\ref{4.8}) of the form $r^\alpha$  one finds
\begin{eqnarray}
\ &&\quad\quad \quad \sum_{k=0}^{\mp}g_{M-k}
\frac{d^{k}r^{\alpha +k-M}\ }{dr^{k}}=r^{\alpha-M}\times 
\label{4.8a}\\
\ &&
\sum_{k=0}^{\mp}g_{M-k}(\alpha+k-M)
(\alpha+k-1-M)\ldots(\alpha+1-M)=0.\nonumber
\end{eqnarray}
Using the descending Pochhammer symbol $(x)_n$, defined as  $(x)_n\equiv x(x-1)(x-2)
\ldots(x-n+1)$ if $n\ge 1$ and $(x)_n\equiv 1$ if $n=0$, equality  (\ref{4.8a}) 
becomes
\begeq\label{4.8b}
\sum_{k=0}^{\mp}g_{M-k}(\alpha+k-M)_k=0.
\endeq
It is a polynomial equation of degree $\mp$ in unknown $\alpha$.
For simplicity one assumes that the roots $\alpha_1,\ldots,\alpha_{\mp}$
are distinct. (The case where some roots coincide will be discussed later.) The
general integral of Equation. (\ref{4.8}) is
\begeq\label{4.9}
Y(r,c_1,\ldots,c_{\mp})\equiv\sum_{k=1}^{\mp} c_k\,r^{\alpha_k}
\endeq
where $c_1,\ldots,c_{\mp}$ are arbitrary constants. \\
A particular integral of the non-homogeneous differential equation
(\ref{4.7}) is easily obtained by the Cauchy method  [see
\S 39 of  Goursat (1959)]. It reads
\begeq\label{4.10}
X(r)=-\int_{r_0}^r Y\left(r,C_1(y),\ldots,C_{\mp}(y)\right)
\left(\frac{y^{m+1}G^{(M+1)}(y)}{g_{m+1}}\right)dy
\endeq                     
where $r_0$ is an arbitrarily chosen value and functions $C_1(y),\ldots,C_{\mp}(y)$
are the solutions of the following system of $\mp$ linear equations
\begin{eqnarray}
Y(y,C_{1},\ldots,C_{\mp}) & = &0,     \nonumber\\
\frac{\partial Y(y,C_{1},\ldots,C_{\mp})} {\partial y}&=&0,   \nonumber\\
                      \ldots &=&0,   \label{4.11}\\
\frac{\partial^{\mp-2}Y(y,C_{1},\ldots,C_{\mp})} {\partial y^{\mp-2}}&=&0,\nonumber\\
\frac{\partial^{\mp-1}Y(y,C_{1},\ldots,C_{\mp})} {\partial y^{\mp-1}}&=&1,\nonumber
\end{eqnarray}                     
that reduces to $Y(y,C_{1},\ldots,C_{\mp}) = 1$ if $\mp=1$.
The $C_i(y)$s solutions of this system are
\begeq\label{4.11a}
C_i(y)=\frac{y^{\mp-1-\alpha_i}}{\prod_{j\ne i}(\alpha_i-\alpha_j)},\quad i=1,\ldots,\mp,
\endeq
and function $ Y\left(r,C_1(y),\ldots,C_{\mp}(y)\right)$ reads
\begeq\label{4.11b}
 Y\left(r,C_1(y),\ldots,C_{\mp}(y)\right)=\sum_{i=1}^{\mp}
\frac{y^{\mp-\alpha_i-1}\,r^{\alpha_i}}{A_i(\alpha)},
\endeq
where it has been put
\begeq\label{4.11c}
A_i(\alpha)\equiv \prod_{j\ne i,\, j=1}^{\mp}(\alpha_i-\alpha_j)\quad {\rm if}\  \ \mp>1,
\endeq
and $A_i(\alpha)\equiv 1$ if $\mp=1$.
The proof  that function $X(r)$, defined by Equation. (\ref{4.10}), is a solution of differential equation (\ref{4.7}) and that (\ref{4.11b}) is a solution of (\ref{4.11})  is reported in appendix A. \\
The general integral of Equation. (\ref{4.7}) is $X(r)+Y(r,c_1,\ldots,c_{\mp})$ and the sought
for  particle probability density has the form
\begeq\nonumber
p(\ell)=\ell^{-3}\left(Y(\ell,c_1,\ldots,c_{\mp})+X(\ell)\right).
\endeq
Condition {\em iii)} is fulfilled if one chooses $r_0=\infty$ and $c_1=c_2=,\ldots,=c_{\mp}=0.$
 In this way the sought for integral transform that allows one to
 obtain the particle size probability density from $G(r)$ is fully determined and reads
\begeq
p(\ell)=\sum_{i=1}^{\mp}\frac{\ell^{\alpha_i-3}}{g_{m+1}A_i(\alpha)}\int_{\ell}^{\infty} y^{M-\alpha_i-1}
\,G^{(M+1)}(y)dy.\quad    \label{4.12}
\endeq  
Finally, the case of degenerate roots can rigorously be dealt with by the 
procedure described in \S 3.3 and 3.4 of Bender \& Orszag (1978). 
In practice one can also proceeds, more simply, as follows. Let  
$\alpha_1=\alpha_2=\ldots=\alpha_r=\alpha$ be a group of $r$ 
coinciding roots. One removes the degeneracy  by  the substitutions  
$\alpha_i\to(\alpha+(i-1)\epsilon)$ with $i=1,\ldots,r$ and
$\epsilon$ equal to a very small number and then apply  the 
described procedure for determining $p(r)$.  By continuity 
the resulting error will be small with $\epsilon$ because all the above 
reported expressions continuously depend on the $\alpha_j$s.
Hereafter one will confine himself to the non degenerate case.    

                           \section{The resolvent kernel in reciprocal space}

The integral transform that determines $p(r)$ by the observed scattering 
intensity $I(q)$ is immediately obtained by taking the $(M+1)$ derivative of
 the rhs of (\ref{4.1a}) and substituting the result into (\ref{4.12}). The result is  
\begeq\label{4.12a}
p(r)=\sum_{i=1}^{\mp}\frac{r^{\alpha_i-3}}{2\pi^2\cC g_{m+1}A_i(\alpha)}
\int_{r}^{\infty} y^{M-\alpha_i-1}
\Bigl[{D_y}^{M+1}\int_0^{\infty}q\sin(qy)I(q)dq\Bigr] dy,  
\endeq 
where $D_y$ denotes the derivative operator with respect to $y$. 
This result  generalizes the Fedorova Schmidt formula because it applies to any CF that  
has a polynomial form as it happens in the case of spheres. It is noted that the 
reconstruction of $p(r)$ from the observed intensity, under the assumption of a 
polynomial CF, holds also true in the case of the slit collimation thanks to 
Guinier's integral transform (Guinier, 1946) that converts $J(q)$, the intensity collected  with the slit geometry, into the pin-hole $I(q)$. \\ 
Our task now is that of rewriting (\ref{4.12a})  in a more compact form that is more convenient for numerical evaluation. To this aim one formally exchanges the integration order in (\ref{4.12a}) and, after putting 
\begeq\label{5.1aa}
K(r,q,\alpha,M)\equiv \int_{r}^{\infty}\,y^{M-\alpha-1}\Bigl[{D_y}^{M+1}
\frac{\sin(qy)}{y}\Bigr]\, dy,
\endeq
 one gets 
\begeq\label{5.1bb}
p(r)= \sum_{i=1}^{\mp}\frac{r^{\alpha_i-3}}{2\pi^2\, \cC\,g_{m+1}A_i(\alpha)}
\int_{0}^{\infty} q\,I(q)\,K(r,q,\alpha_i,M)\, dq.  
\endeq
It is noted that definition (\ref{5.1aa}) does not always lead to an existing $K(r,q,\alpha,M)$ because, according to  Riemann's theorem (Bender \& Orszag, 1978), 
the integral exists only if $M-2-\alpha<0$, a condition  that is 
not {\em a priori} ensured. However each integral contributing to equation (\ref{4.12}) exists owing to assumption {\em iii)}.  Consequently the convergence difficulty of  (\ref{5.1aa}) is related to the fact that the integration order cannot be  exchanged 
for the $\alpha_i$s such that $\alpha_i<(M-2)$. Hence one must  handle 
the case $\alpha_i>(M-2)$ differently from that where $\alpha_i<(M-2)$. This will be 
respectively done in the following  two subsections.

                   \subsection{ The case $\alpha_i>(M-2)$ }  
Assume, for simplicity, that all the $\alpha_i$s are greater than $(M-2)$. In this case 
the order of integration is not important and the corresponding $K(r,q,\alpha_i,M)$s 
do exist. These can be expressed in terms of some known transcendental functions proceeding as follows. Perform the integration variable change $y\to rt$ in (\ref{5.1aa}) and momentarily omit  index $i$ for notational simplicity. 
One gets
\begeq\label{5.1.1}
K(r,q,\alpha,M)=
r^{-\alpha-2}\int_1^{\infty}t^{M-\alpha-1}\frac{d^{M+1}\  }{dt^{M+1}}\left(\frac{\sin(q\,r\, t)}{t}\right)dt.
\endeq 
Putting $Q_r\equiv q\,r$, using the identity [Luke, \S\,2.9, (1969)]
\begeq\label{5.1.2}
\frac{d^{m}\left(\sin(qy)/y\right)}{dy^{m}} = \frac{q^{m+1}}{y^{m+1}}\frac{d^{m}\left(\sin(qy)/q\right)}{dq^{m}}, \ \
 m=0,1,2,\ldots,
\endeq
 and denoting the derivation operator with respect to $Q_r$ by $D_{Q_r}$, the previous integral becomes
\begeq
r^{-\alpha-2}{Q_r}^{M+2}D_{Q_r}^{M+1}\int_1^{\infty}t^{-\alpha-3}\frac{\sin(Q_r t)}{Q_r}dt.\nonumber
\endeq 
Recalling the definition of $S(x,a)$, the generalized Fresnel sine integral (equation  (6.5.8) of Abramowitz \& Stegun, 1970), one finds that
\begeq\label{5.1.3}
\kappa_A(Q_r,\alpha)\equiv 
\int_1^{\infty}t^{-\alpha-3}\frac{\sin(Q_r t)}{Q_r}dt = Q_r^{\alpha+1}S(Q_r,-2-\alpha),
\endeq
which can also be expressed  in terms of the hypergeometric function $_1F_2(a;b,c;x)$ 
(Luke, 1969) as 
\begeq\label{5.1.4} 
\kappa_A(Q_r,\alpha)=
{Q_r}^{1+\alpha}
\Gamma(-2-\alpha)\sin\frac{\pi\alpha}{2}+\frac{1}{1+\alpha} { _1F_2}\left(-\frac{1+\alpha}{2}; \frac{3}{2},\frac{1-\alpha}{2}; -\frac{Q_r^2}{4}\right).
\endeq
Combining equations (\ref{5.1bb}), (\ref{5.1.1}),  (\ref{5.1.2})  and  (\ref{5.1.3}) one finds 
\begeq\label{5.1.6}
p(r)= \frac{r^{-5}}{2\pi^2\, \cC\,g_{m+1}}
\int_{0}^{\infty} q\,I(q)\cK_A(q\,r)dq
\endeq
with 
\begeq\label{5.1.7}
\cK_A(Q_r)\equiv {Q_r}^{M+2}D_{Q_r}^{M+1}\sum_{i=1}^{\mp}\frac{1}{A_i(\alpha)}\kappa_A(Q_r,\alpha_i).
\endeq  
For later convenience it is also mentioned that the leading asymptotic expansion 
of $\kappa_A(q,\alpha)$ at large $Q_r$s is 
\begin{eqnarray}\label{5.1.5}
\kappa_A(Q_r,\alpha)&\approx& \frac{\cos Q_r}{{Q_r}^2}\left(1-\frac{12+7\alpha}{{Q_r}^2}+
\ldots\right) +\nonumber\\
\quad&&\quad \frac{\sin Q_r}{{Q_r}^3}\left(3+\alpha-\alpha\frac{47+12\alpha+\alpha^2}{{Q_r}^2}+
\ldots\right).
\end{eqnarray}
                  \subsection{ The case $\alpha_i<(M-2)$ } 
One assumes that  all the  $\alpha_i$s are smaller than $(M-2)$. In this case, as it was already said, the convergence at $\infty$ of each  integral 
present in (\ref{4.12}) is ensured by the sufficiently fast decrease of $G^{(M+1)}(y)$. Hence, one writes   
\begeq\label{5.2.1}
\int_{r}^{\infty} y^{M-\alpha_i-1}\,G^{(M+1)}(y)dy=\cGh_i-\int_{0}^{r} y^{M-\alpha_i-1}\,G^{(M+1)}(y)dy
\endeq
with 
\begeq\label{5.2.2}
\cGh_i\equiv \lim_{r\to 0} \int_{r}^{\infty} y^{M-\alpha_i-1}\,G^{(M+1)}(y)dy.
\endeq 
Owing to condition {\em iii)}, the limit of $p(r)$ as $r\to 0$ cannot be divergent. Then, from equation (\ref{4.12}), 
it  follows that $\cGh_i=0$ whatever $i$. 
Expressing $G^{(M+1)}(y)$  in terms of its FT and omitting index $i$ for simplicity, one finds 
\begeq\nonumber
-\int_{0}^{r} y^{M-\alpha-1}\,G^{(M+1)}(y)dy=-\int_{0}^{r} dy \int_0^{\infty}q\, dq\,y^{M-\alpha-1}
\frac{{D_y}^{(M+1)}}{2\pi^2 q\,\cC } 
\left(\frac{\sin(q y)}{y}\right)\,I(q).
\endeq
The order of integration can now be exchanged. Using again identity (\ref{5.1.2})  
the last integral becomes 
\begeq\label{5.2.4}
\frac{1}{{2\pi^2 q\,\cC }}\int_{0}^{\infty} q\,I(q)\, dq \left(-\int_0^{r} dy\,{y^{-\alpha-3}}{q^{M+2}}
{{D_q}^{(M+1)}} 
\left(\frac{\sin(q y)}{q}\right)\right).
\endeq
A further change of the integration variable converts the above $y$-integral into 
\begeq\label{5.2.5}  
r^{-\alpha-2}{Q_r}^{M+2}{D_{Q_r}}^{(M+1)}\left(-\int_0^1 t^{-\alpha-3}\left(\frac{\sin(Q_r y)}{Q_r}\right)dt\right).
\endeq
This integral also can be expressed in terms of an $_1F_2(.)$ hypergeometric function, since
\begin{eqnarray}
\kappa_B(Q_r,\alpha)&\equiv& -\int_0^1 t^{-\alpha-3}\left(\frac{\sin(Q_r y)}{Q_r}\right)dt=\label{5.2.6}\\
\quad&&\frac{1}{1+\alpha}{ _1F_2}\left(-\frac{1+\alpha}{2}; \frac{3}{2},\frac{1-\alpha}{2}; -\frac{Q_r^2}{4}\right),
\nonumber
\end{eqnarray}
\ie\ the opposite of the hypergeometric contribution present in the rhs of (\ref{5.1.4}). It follows that, at large 
$Q_r$, the leading asymptotic term of $\kappa_B(Q_r,\alpha)$ is the opposite of that of $\kappa_A(Q_r,\alpha)$ 
[see (\ref{5.1.5})] plus ${Q_r}^{1+\alpha}\Gamma(-2-\alpha)\sin\frac{\pi\alpha}{2}$.\\ 
Combining equations (\ref{5.2.1})-(\ref{5.2.6}) and substituting the result in (\ref{4.12a}) one finds that 
\begeq\label{5.2.7}
p(r)= \frac{r^{-5}}{2\pi^2\, \cC\,g_{m+1}}
\int_{0}^{\infty} q\,I(q)\cK_B(q\,r)dq
\endeq
with 
\begeq\label{5.2.8}
\cK_B(Q_r)\equiv {Q_r}^{M+2}D_{Q_r}^{M+1}\sum_{i=1}^{\mp}\frac{1}{A_i(\alpha)}\kappa_B(Q_r,\alpha_i).
\endeq
        \subsection{The general case of some $\alpha_i$s smaller and the others greater  than $M-2$}
The expression of the integral transform that determines the particle size distribution from the scattering intensity, whatever the $\alpha_i$ values,  immediately follows from (\ref{5.1.7}) and (\ref{5.2.8}). 
It is 
\begeq\label{5.fnla}
p(r)= \frac{r^{-5}}{2\pi^2\, \cC\,g_{m+1}}
\int_{0}^{\infty} q\,I(q)\cK_G(q\,r)dq
\endeq
with 
\begeq\label{5.fnlb}
\cK_G(Q_r)\equiv {Q_r}^{M+2}D_{Q_r}^{M+1}\biggl[{\sum_{i=1}^{\mp}}\, '\frac{1}{A_i(\alpha)}\kappa_A(Q_r,\alpha_i)  +   
{\sum_{i=1}^{\mp}}\,''\frac{1}{A_i(\alpha)}\kappa_B(Q_r,\alpha_i)\biggr],
\endeq
where the $'$ and the $''$ respectively denote that the corresponding sums are restricted to 
$\alpha_i$s greater and smaller than $(M-2)$, and $Q_r=q\,r$.  \\ 
Owing to assumption {\em iii)} the behaviour of the general resolvent kernel $\cK_G(q\,r)$ 
is such as to ensure the convergence of (\ref{5.fnla}) 
even though the property does not apply to each term of the sums. Further, 
the convergence is rather weak as it will appear clear from the discussions reported in the following section.      
             \section{Polynomial approximation of the CF} 
At this point one wonders: besides the spherical shape, do exist other particle shapes 
that have a polynomial CF?  As yet the answer is unknown. But this does not imply 
that one cannot use the above finding to approximate the CF of particles with a given 
shape by a polynomial. To construct such approximations one first recalls  that $\gamma(r)$, the CF  of a unit particle with a given shape, obeys some general constraints. The simplest of these  are:
\begin{itemize}
\item {\em a)} $\gamma(0)=1$,
\item {\em b)} $\gamma\p(0)=\sigma$ with $\sigma\equiv -s/4v$ ($\sigma$ will be hereafter referred to as {\em specific surface}),
\item {\em c)} $\gamma(1)=0$,
\item {\em d)} $\gamma\p(1)=0$,
\item {\em e)} \ 4$\pi\int_0^{\infty} r^2\gamma(r)dr=V$,
\item {\em f)} \ 4$\pi\int_0^{\infty}  r^4\gamma(r)dr=2V{R_G}^2$ where $R_G$ denotes
the Guinier gyration radius of the particle,
\item {\em g)} $\gamma\pp(0)=\cA$ where $\cA$ is the particle {\em angularity} defined by equation (\ref{2.4b}) (Porod, 1967;  M\'ering and Tchoubar, 1968;   Ciccariello \etal; 1981),  and
\item  {\em h)} $\gamma\ppp(0)=\cK$ where $\cK$, defined by equation (\ref{2.4c}), is the particle {\em curvosity} (Kirste \& Porod. 1962; Ciccariello \& Sobry, 1995).
\end{itemize}
Besides the listed constraints one might consider other ones.  For instance, if the particle
shape is such that one does not have a parallelism condition between
finite area subsets of the particle surface at a relative distance equal to the particle
maximal chord, one also has $\gamma\pp(1)=0$. Similarly, {\em e)} and {\em f)}
are particular cases of the general relation
\begin{eqnarray} \label{6.1a}
\cG_{2m}&\equiv&\ 4\pi\int_0^{\infty} r^{2m}\,r^2\gamma(r)dr=V{\sum\,'}_{0\le h,k,l\le m}\frac{m!a_l}
{h! k! l!} \times \\
 &&\quad \langle{R^{2h+l}}\rangle \langle{R^{2k+l}}\rangle,\quad m=0,1,2,\ldots\nonumber
\end{eqnarray}
that connects the $2m$th moment of the particle CF to the higher order gyration radii 
of the particle (see Appendix B for the proof of  (\ref{6.1a}) and for the explanation of  
the symbols there involved).\\
According to Shannon's theorem the information content
of any SAS intensity is not particularly rich so that a ten of parameters  only  can be determined 
(Moore,  1980;  and Taupin \& Luzzati, 1982). Thus, in performing a polidisperse
analysis of the SAS intensity of a given sample, it appears sensible to approximate
the particle CF by a polynomial that fulfills  some of the above constraints. \\
In this paper the simplest cases of the third [$P_3(r)$] and fourth  [$P_4(r)$] degree polynomial approximations  will be considered. In particular, it will be required that the polynomial 
approximation obeys  {\em a),  b),  c)}  and {\em d)} that explicitly accounts for  the support 
properties of the CF, the normalization at $r=0$ and a single geometrical feature of the
 particle, namely  its specific surface $\sigma$.  For the 3rd degree case one  finds that
\begeq\label{6.1b}
\gamma(r)\,\approx\, P_3(r)\, =\, 1 + \sigma\,r  - (3 + 2 \sigma) r^2  + (2 + \sigma) r^3.
\endeq
The associated angularity, curvosity, volume and Guinier gyration radius respectively are
\begin{eqnarray}\label{6.1c}
\cA =   - (3 + 2 s),&\quad& \cK=6(2+\sigma),\\
\quad V= \pi (4 + \sigma)/15,&\quad&
{R_G}^2=(18 + 5 \sigma)/(28 (4 + \sigma)).\nonumber
\end{eqnarray}
From (\ref{6.1b}) follows that the $g_m$ coefficients, defined by (\ref{4.3a}), are
\begeq\label{6.1d}
g_0=g_1=0,\   g_2=g_2(\sigma)=2 (3 + \sigma),\  g_3=g_3(\sigma)=6 (2 + \sigma).
\endeq
One clearly has $M=3$, $m=1$, $\mp=1$ and the solution of  polynomial
equation (\ref{4.8b}) is
\begeq\label{6.1f}
\alpha=\alpha(\sigma)=-\sigma/(3 + \sigma).
\endeq 
\begin{figure}[!h]
{\includegraphics[width=7.truecm]{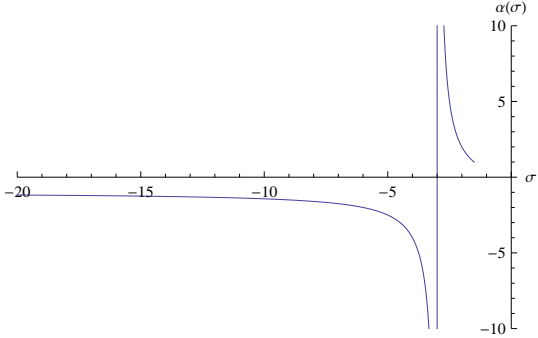}}
\caption{\label{Fig1} {Behaviour of the root of Equation. (\ref{4.8b}) in terms of the 
specific surface parameter $\sigma$ when one considers the 3rd degree polynomial 
approximation of the particle CF. }}
\end{figure}
Figure 1 plots $\alpha(\sigma)$ {\em vs.} $\sigma$. It is recalled that $\sigma$ must be such  that $\sigma<-3/2$ because,  at fixed volume, the sphere is the geometric solid
that has the smallest surface. Thus, $\alpha$  varies between $-\infty$ and
$\infty$ and the variation is very sharp around $\sigma=-3$.\\ 
In the following subsections we shall  analyze the case of the 
sphere, that of the cube and the octahedron,   that of the tetrahedron and, finally, that of the 
4th degree approximation.
               \subsection{The sphere case} 
This case, already fully exploited, is mainly  reported in order to make fully 
evident that equations (\ref{5.fnla}) and (\ref{5.fnlb}) coincide with those 
of Fedorova and Schmidt.
As it was just said the specific surface of the sphere is $\sigma= -3/2$. The substitution
of this value in (\ref{6.1b}) reproduces the exact CF of the unit sphere, \ie
\begeq\label{6.1.1}
\gamma_{sph}(r)=1-3r/2+r^2/4.
\endeq 
This nice property implies that the  constraints relevant to the angularity, the curvosity and all the Guinier  higher order gyration radii [see   (\ref{6.1c})] 
are exactly obeyed. \\
As already anticipated it is now shown that integral transform (\ref{4.12}) coincides with
Fedorova and Schmidt's one  in the case of spherical particles. From Eqs. (\ref{6.1d})
and (\ref{6.1f}) it  follows that $g_2=3$ and $\alpha=1$. Hence  Equation. (\ref{4.12}) reads
\begin{equation}\label{6.1.2}
p(r)=\frac{1}{3r^2}\int_r^{\infty} yG^{(4)}(y)dy.
\end{equation}
Integrating twice by parts, one obtains
\begeq\label{6.1.3}
p(r)=\frac{1}{3r^2}\left(-rG^{(3)}(r)+G^{(2)}(r)\right)=-\frac{1}{3}
\frac{d\ }{dr}\left(\frac{G^{(2)}(r)}{r}\right),
\endeq
that coincides with  the result of Letcher and  Schmidt (1966) and Fedorova
\& Schmidt (1978). \\  
It is instructive to check equations (\ref{4.12}) and (\ref{4.12a})
choosing as  size probability density  the  $(n,\lambda)$ Poisson one, namely
\begeq\label{6.1.4}
p(n.\lambda,r) \equiv r^n\,e^{-\lambda r}/n!,
\endeq
with $n=4$ and $\lambda=1$. In the sphere case, the explicit evaluation of  ({\ref{4.1}) yields 
\begeq\label{6.1.5}
G(r)={ e^{-r}}\left (1680 + 1320 r + 480 r^2 + 104 r^3 + 14 r^4 + r^5\right)/{8}.
\endeq
Substituting this expression in the rhs of (\ref{6.1.3}) one straightforwardly
verifies that the result coincides with  $p(4,1,r)$. \\
The check of  (\ref{4.12a}) is more interesting. Since $\alpha=1$,  the size distribution 
is determined by   (\ref{5.1.6}) and (\ref{5.1.7}). The resolvent kernel $\cK_A(q\, r)$,  now 
denoted  as $\cK_{sph}(q\,r)$, is 
\begeq\label{6.1.6}
\cK_{sph}(q\,r)= (q\,r)^3\kappa_{sph}(q\,r), 
\endeq 
with
\begeq\label{6.1.6a}
\kappa_{sph}(q\,r)\equiv \cos(q\,r) \bigl[1-\frac{8} {(q\,r)^2}\bigr ] - 
4\, \frac{\sin(q\,r)}{q\,r}\bigl[1-\frac{2} {(q\,r)^2}\bigr] 
\endeq
and (\ref{5.1.6}) becomes
\begeq\label{6.1.7}
p(r)=\frac{1}{6\,\pi^2\,\cC\,r^{2}}\int_0^{\infty}q^4I(q)\,\kappa_{sph}(q\,r)\,dq.
\endeq
Function $\kappa_{sph}(q\,r)$ is such that the integral $\int_0^{\cQ_M}\kappa_{sph}(q\,r)dq$ can be set equal to zero as $\cQ_m\to\infty$ because at large $\cQ_M$ it  
behaves as $\sin(\cQ_M\,r)/r$ that, it being  wildly oscillating if $r\ne 0$,  averages to 
zero.   Hence, equation (\ref{6.1.7}) can be written in the well known form  
\begeq\label{6.1.8}
p(r)=\frac{1}{6\,\pi^2\,\cC\,r^2}\int_0^{\infty}\big(q^4I(q)-\prd )\,\kappa_{sph}(q\,r)\,dq,
\endeq
which  converges faster than (\ref{6.1.7}) at $q=\infty$.  Its correctness can explicitly be checked  in the case of  a (4,1) Poisson distributed spherical particles. 
The explicit form $ I_{p,sph}(q)$ of $I(q)$ is obtained by (\ref{3.9}) and reads
\begeq\label{6.1.9}
 I_{p,sph}(q)=24\,\cC \,\frac{ 1050 + 420 q^2 + 567 q^4 + 329 q^6 + 107 q^8 + 15 q^{10}}{(1 + q^2)^7}.
\endeq
From this expression one finds that  $\prd=360\,\cC$.  Substituting these expressions in the rhs of (\ref{6.1.8})  and evaluating the integral one correctly  finds function $p(4,1,r)$. 
It is stressed that (\ref{6.1.8})  numerically is much more convenient than (\ref{4.12a}) to 
get $p(r)$ from the observed scattering intensity. [A more convenient form 
has been recently discussed by Botet \& Cabane (2012).] Finally, figure 2 shows the size 
distribution obtained by the numerical evaluation of (\ref{6.1.8}) using as $I(q)$ 
the values resulting from the numerical integration of (\ref{3.9}) in the sphere case  (with $\cC=1$). 
\begin{figure}[!h]
{
\includegraphics[width=7.truecm]{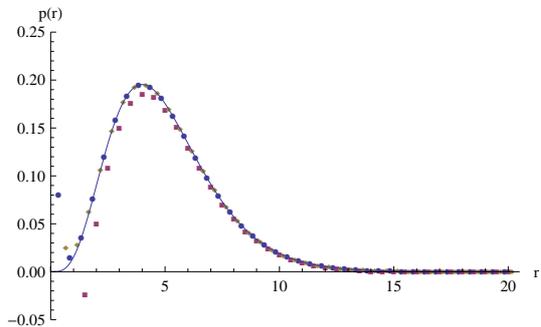}
}
\caption{\label{Fig2} {The thin curve plots the (4,1) Poisson distribution.
The magenta full squares are the values obtained by Equation. (\ref{6.1.8}) and a numerical
evaluation of (\ref{3.9}) performed as explained at the end of \S\,6.2. The 
blue full circles and the golden diamonds respectively are the values of $p(r)$ 
reconstructed in the octahedron/cube by (\ref{6.2.1.3}) and in the tetrahedron case by (\ref{6.2.2.3}) (see the end of \S\,6.2).}}
\end{figure}

\subsection{The 3rd degree polynomial approximation of the cube's,   octahedron's and tetrahedron's CFs }
Ciccariello and Sobry (1995) showed that the CF of any polyhedral particle is a 3rd degree
polynomial in the innermost $r$-range.   This condition clearly is obeyed by the known 
CFs of the regular tetrahedron, octahedron (Ciccariello, 2014) and cube (Goodisman, 1980). In these cases the innermost $r$-range is larger than half  the total $r$-range where the CFs differ from zero. For this reason it  is tempting to approximate the known CFs by a 3rd degree polynomial, vanishing together with its first derivative at the outermost $r$ value and subsequently  perform a polidisperse analysis along the lines expounded in §6.1. 
The results of this approximation will now be  illustrated.\\
One knows that the specific surface values  of the unit cube, the unit octahedron and the unit tetrahedron  are respectively equal to 
\begeq\label{6.2.1}
\sigma_{C}=-3 \sqrt{3}/2 ,\quad \sigma_{O}= -3 \sqrt{3}/2,\quad{\rm and}\quad \sigma_{T}=-3 \sqrt{3/2}.
\endeq 
The 3rd degree polynomial approximations of the three CF immediately result from 
the substitution of the above $\sigma$ values into (\ref{6.1b}). 
Since the $\sigma$ values of the cube and the octahedron coincide the resulting 3rd degree approximation of the cube CF is equal to that of the octahedron. \\ 
\begin{figure}[!h]
{
\includegraphics[width=7.truecm]{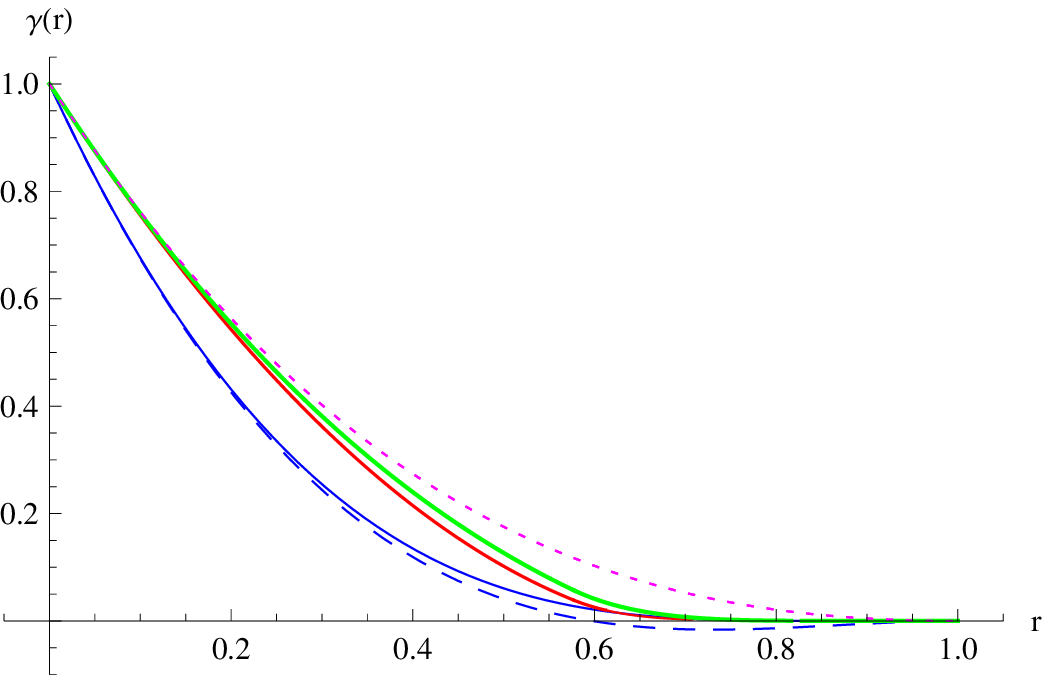} \includegraphics[width=7.truecm]{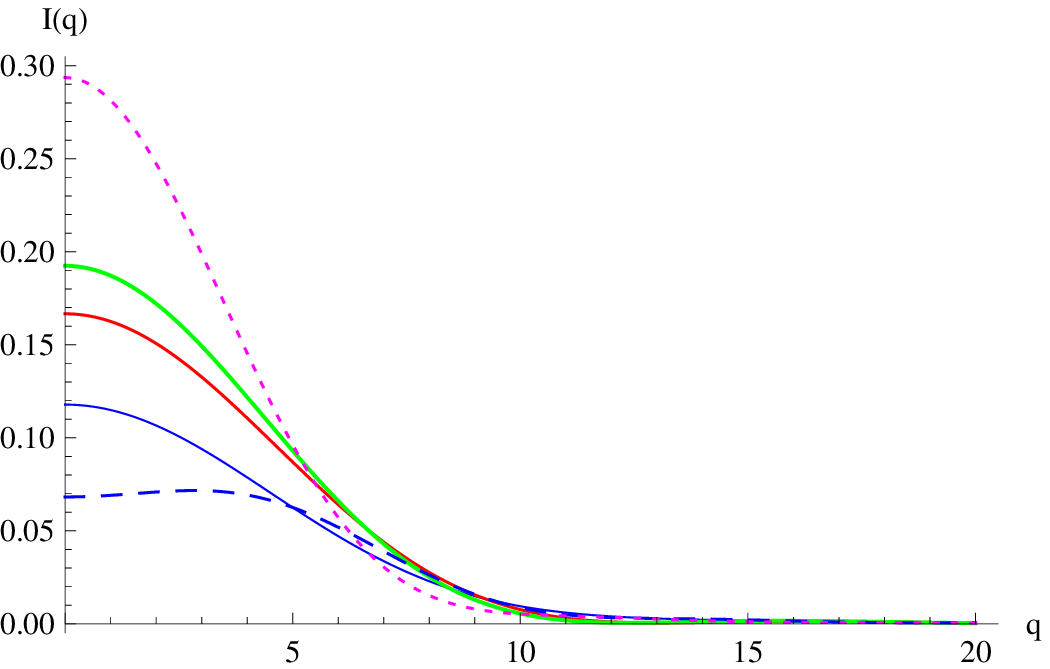} }
\caption{\label{Fig3} {Left: Comparison of the exact CFs with those obtained by the 3rd 
degree polynomial approximations described in the text. The continuous and broken blue 
curves refer to the exact and the polynomial approximation of the tetrahedron CF.   
The continuous  red and green ones to the exact CFs of the octahedron and the 
cube while  the dotted magenta curve plots their 3rd degree approximation. Right: 
Behaviour of the FTs  of the CFs shown in the left 
panel. The symbols are the same.   }}
\end{figure} 
Figure 3 plots the exact CFs and their 3rd degree approximations as well as the FTs 
of  the exact and the polynomial approximated CFs.   In direct space the agreement 
is relatively good for the tetrahedron, reasonable for the cube and not bad for the 
octahedron since small discrepancies are only present in the outermost $r$/range. 
These discrepancies are responsible for those observed at small $q$'s  in reciprocal 
space.  They can be reduced requiring that the polynomial 
approximations also obey  
constraint {\em e)} because the fulfillment of this constraint implies that the 
FTs of the exact and the polynomial approximated CFs coincide at $q=0$. 
To do that one must consider  a 4th degree polynomial approximation, a case 
discussed  in \S 6.3.\\   
We shall go on  with the discussion of the 3rd degree 
polynomial approximation  in the cases of the (4,1) Poisson distributions of cubes, octahedra 
or tetrahedra. in order to make clear all the point of the analysis.  \\
By Equation. (\ref{6.1b}) one finds that function $G(r)$, defined by Equation. (\ref{4.2}), becomes 
\begin{eqnarray}\label{6.1x}
G_3(r,\sigma)&=&e^{-r }\bigl[210 + 30 r (7 + \sigma) + 10\,r^2 (9 + 2 \sigma)  +\\ 
& & 2 r^3\,(11 + 3 \sigma)+   r^4\,  (13 + 4 \sigma)/4 + r^5 (3 + \sigma)/12\bigr].\nonumber
\end{eqnarray}
Its 3D FT, multiplied by $\cC$,  yields the polidisperse scattering intensity, \ie  
\begin{eqnarray}\label{6.2x}
I_3(q,\sigma)&=&\frac{48 \pi\,\cC} {(1 + q^2)^7}\bigl [210 (4 + \sigma) - 60 q^2 (1 + 3 \sigma) +\\ 
& &  7 q^4 (12 - 19 \sigma) + 7 q^6 (4 - 13 \sigma) + q^8 (4 - 13 \sigma) - 5 q^{10} \sigma)\bigr].\nonumber
\end{eqnarray}
Substituting in the above two relations the $\sigma$ values reported in Equation.(\ref{6.2.1})  one 
obtains the CFs as well as  the scattering intensities relevant to the three collections  of $p(4,1,d)$ 
Poisson polidisperse tetrahedrons, octahedrons and cubes.  The corresponding exact values are 
obtained, using the exact particle CFs,  by a numerical evaluation of  integrals (\ref{4.2}) and  (\ref{3.9}), which  can more conveniently be evaluated by  
\begeq\label{6.3x}
I_p(q)= \int_1^{\infty} y^3\,\gamma(1/y)\,{\tilde p}(4,1,q,y) dy,
\endeq
with        
\begin{eqnarray}\label{6.4x}
&&{\tilde p} (4,1,\,q,\,y) \equiv \frac{4\pi } {q}\int_0^{\infty}r^5\,\sin(q\,r)p(4,1,r\,y)dr=\\
&&\quad\quad \frac{3\,\pi(8!)\, q\, y^5\, (5\, q^4 - 10\, q^2\, y^2 + y^4) (q^4 - 10\, q^2 y^2 + 5\, y^4)}
{(q^2 + y^2)^{10}}.\nonumber
\end{eqnarray}
\begin{figure}[!h]
{
\includegraphics[width=7.truecm]{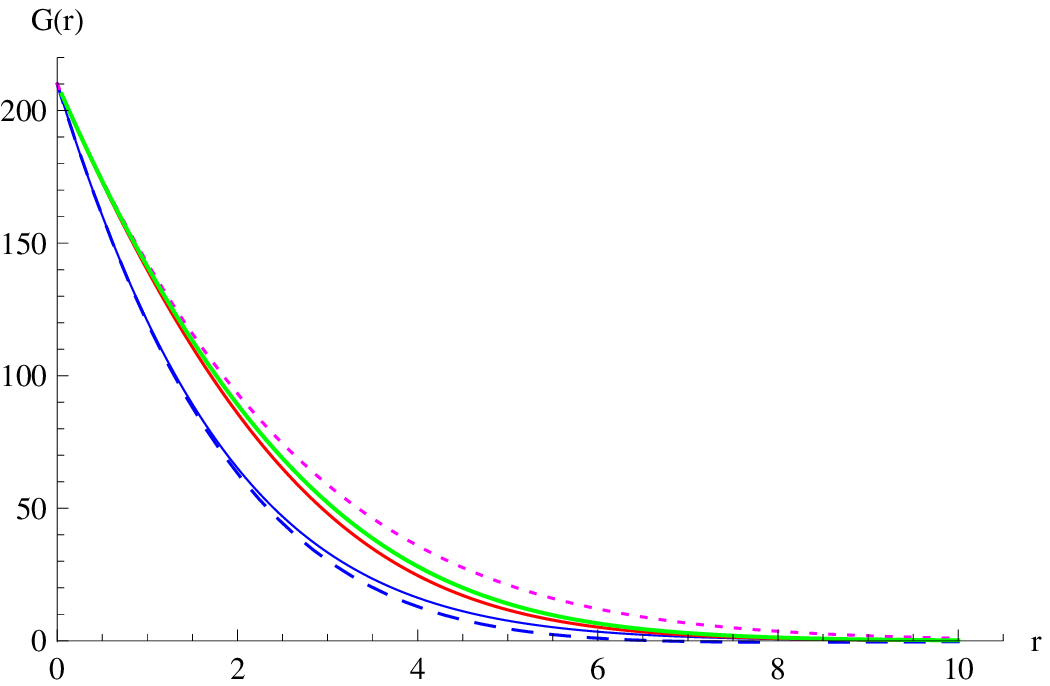}\includegraphics[width=7.truecm]{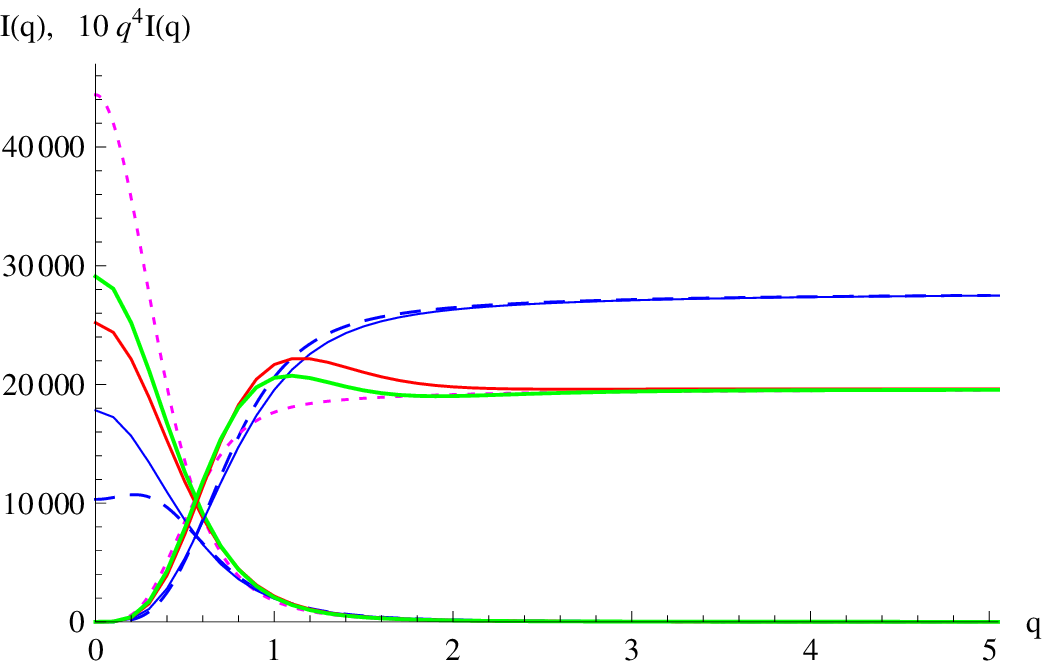} }
\caption{\label{Fig4} {Using the same conventions of Fig. 3, the left panel 
shows the exact (continuous curves) and the approximated (broken curves) CFs relevant 
to the (4,1) Poisson polidisperse samples of tetrahedra, octahedra and cubes. 
The right panel shows the intensities as well as their Porod plots.    }}
\end{figure} 
The results are shown in Figure. 4. The left panel shows the $G(r)$'s and the right 
one the intensities and their Porod plots.  For the intensities, the quality of the agreement is similar to that shown in Fig. 3. However, the Porod plots show that the intensity discrepancies observed near the origin are washed out by the factor $q^4$.  Some discrepancies are still observed around $q=1$ in the only cubic and octahedral cases,  while the agreement is quite satisfactory in the tetrahedron case. \\ 
\subsubsection{Determination of the size distribution from the scattering intensity 
 in the cube/octahedron case} 
 The root of the resolvent equation relevant to the 3rd degree polynomial approximation of  a  particle CF is given by equation (\ref{6.1f}). Since the specific surfaces of the unit octahedron and the unit cube coincide it follows that $\alpha$ is
the same in the two case. It will be denoted by $\alpha_{o/c}$. Recalling that $\sigma$ is equal to $\sigma_{o/c}\equiv -3^{3/2}/2$ [see (\ref{6.2.1})], one finds 
\begeq\label{6.2.1.1}
 \alpha_{o/c}\equiv\alpha(\sigma_{o/c})=3 + 2\, \sqrt{3}\approx 6.46.
\endeq
The resolvent kernel (\ref{5.1.7}) takes now the form  
$\cK_{o/c}(q\,r)=(q\,r)^3\kappa_{o/c}(q\,r)$,   with 
\begin{eqnarray}
\kappa_{c/o}(Q_r)&\equiv&\Bigl(1 - \frac{12  - 5 \alpha_{o/c}+{\alpha^2}_{o/c}}{{Q_r}^2}\Bigr)\cos Q_r+\Bigl(\alpha_{o/c}-5+\label{6.2.1.2} \\ 
 &&\frac{12-7\alpha_{o/c}+ 4{\alpha}^2_{o/c}-{\alpha}^3_{o/c}}{Q_r^2}\Bigr)
\frac{\sin Q_r}{Q_r}+{\alpha}^3_{o/c}\sin\frac{\pi\,{\alpha}_{o/c}}{2}\times\nonumber\\ 
&&\big({\alpha}_{o/c}-2\bigr)\Gamma(-2-{\alpha}_{o/c})+\frac{{\alpha}_{o/c}}{{Q_r}^2}
(2-3{\alpha}_{o/c}+{\alpha^2}_{o/c})\times\nonumber\\ 
 &&{ _1F_2}\left(-\frac{1+\alpha_{o/c}}{2}; \frac{3}{2},\frac{1-\alpha_{o/c}}{2}; -\frac{Q_r^2}{4}\right),
\nonumber
\end{eqnarray}
and $Q_r=q\,r$. Also in this case one finds that $\int_0^{\infty}\kappa_{c/o}(q\,r)dq=0$ 
in the sense reported just below (\ref{6.1.7}), so that (\ref{5.1.6}) can be written as
\begeq\label{6.2.1.3}
p(r)= \frac{1}{2\pi^2\, \cC\,g_2\,r^2}
\int_{0}^{\infty} (q^4\,I(q)-\prd)\,\kappa_{o/c}(q\,r)dq
\endeq 
that is more convenient for numerical computation because the integrand behaves 
as $const\times \sin(q\,r)/q$ at large $q$s. \\ 
The result (\ref{6.2.1.3}) has  been analytically checked substituting, in its rhs,  $I(q)$ with  
$I_3(q,s_{o/c})$ given by expression (\ref{6.2x}) and setting $\prd=\lim_{q\to\infty}I_3(q,s_{o/c})$. The result is the outset Poisson (4,1) size 
distribution.   \\
\subsubsection{Determination of the size distribution from the scattering intensity in the tetrahedron case}
The specific surface of the tetrahedron is given by (\ref{6.2.1}c). Then the root of the 
resolvent equation associated to the polidisperse polynomial approximation of 
tetrahedrons is given by (\ref{6.1f}) and reads 
\begeq\label{6.2.2.1}
 \alpha_{t}\equiv\alpha(\sigma_{T})=-3 - \sqrt{6}\approx-5.45.
\endeq
It is smaller than one  and therefore one must apply the results of \S 5.2.
The resolvent kernel, defined by (\ref{5.2.8}), is $\cK_t(q\,r)=(q\,r)^3\kappa_t(q\,r)$
with
\begin{eqnarray}\label{6.2.2.2}
\kappa_t(q\,r)&\equiv &\cos(q\,r)\Bigl(1-\frac{12-5\alpha_t+{\alpha_t}^2}{q^2r^2}\Bigr)-\\
&&\frac{\sin(q\,r)}{q\,r}\Bigl(5-{\alpha_t}-\frac{12-7{\alpha_t}+4{\alpha_t}^2-{\alpha_t}^3}{q^2r^2}\Bigr)+
\nonumber\\
&& +\frac{\alpha_t\,(2-3{\alpha_t}+{\alpha_t}^2)}{q^2r^2}
{ _1F_2}\left(-\frac{1+\alpha_t}{2}; \frac{3}{2},\frac{1-\alpha_t}{2}; -\frac{q^2r^2}{4}\right).
\nonumber
\end{eqnarray}
$\kappa_t(q\,r)$ also is such that $\int_0^{\infty}\kappa_{r}(q\,r)dq=0$ in the weak 
sense. One concludes that the particle size distribution in the tetrahedron polynomial 
approximation is given by 
\begeq\label{6.2.2.3}
p(r)= \frac{1}{2\pi^2\, \cC\,g_2\,r^2}
\int_{0}^{\infty} (q^4\,I(q)-\prd)\,\kappa_{t}(q\,r)dq.
\endeq
The analytic check of this relation is fully satisfactory as for the case of equation (\ref{6.2.1.3}) .\\ 
\begin{figure}[!h]
{
\includegraphics[width=7.truecm]{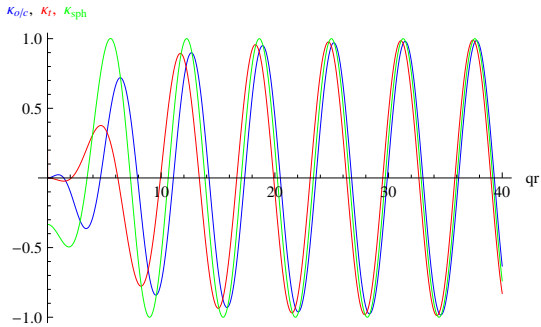} }
\caption{\label{Fig5} {Plots of the resolvent kernels relevant to the sphere case 
[equation  (\ref{6.1.6a}), green curve], to the cube/octahedron case [equation (\ref{6.2.1.2}), blue curve] 
and to tetrahedron case [equation  (\ref{6.2.2.2}), red curve].  }}
\end{figure} \noindent 
\vskip 0.5truecm \noindent  
As in the case of Fig. 2, equations (\ref{6.2.1.3}) and (\ref{6.2.2.3}) have also been numerically checked as follows. First one evaluates the scattering intensity, 
given by equation (\ref{6.2x}),  for the octahedron/cube and the tetrahedron cases 
on a grid of 1000 points uniformly covering the interval $0<q<50$. Then one evaluates 
integrals (\ref{6.2.1.2}}) and (\ref{6.2.2.2}) using an $r$-grid of 40 values uniformly 
distributed over the interval [0,\,20]. The resolvent kernels (including the sphere case) are plotted in figure 5 and the resulting size distributions are shown in Fig. 2. The agreement 
is quite satisfactory since the discrepancies observed at small $r$s are related to 
the $q$ truncation. For a general discussion of this point see Pedersen (1994).\\  
\begin{figure}[!h]
{ 
\includegraphics[width=7.truecm]{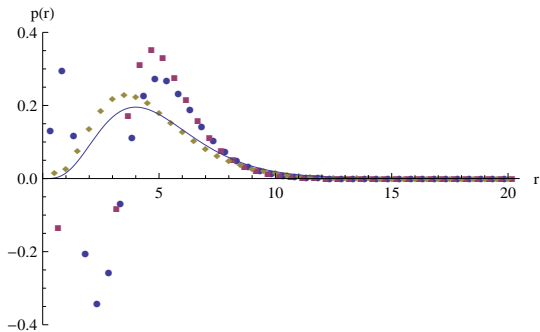} 
}
\caption{\label{Fig6} {Reconstruction of $p(r)$ by (\ref{6.2.1.3}})  using the 'exact'  
scattering intensities  of the  polidisperse octahedrons (blue full circles), cubes 
(magenta full squares) and  tetrahedrons (golden full diamonds). The continuos 
curve is the outset (4,1) Poisson distribution.}
\end{figure} \noindent 
The $p(r)$s have also been reconstructed, along the lines just reported, using as scattering intensities 
those obtained by the exact particle CFs, \ie\, using  equations  
(\ref{6.3x}) and (\ref{6.4x}).  The results are shown in figure 6, where the 
blue full circles, the magenta full squares and the golden full diamonds respectively 
refer to the polidisperse octahedra, cubes and tetrahedra. The figure represents a first 
test on the reliability of approximating an exact CF by a polynomial one to 
perform a polidisperse analysis. It shows that the resulting $p(r)$ is reliable 
if the particle exact scattering  intensity is reasonably approximated by the polynomial 
polidisperse one. It is noted that the agreement must be observed in the Porod plot of the 
intensities because the inversion formulae involve quantity $q^4I(q)$ in their integrands. This appears to be the case of tetrahedral particles, while the 3rd degree polynomial approximation is not equally satisfactory in the case of cubes and octahedra. 
In fact, one sees that the resulting $p(r)$s are satisfactory in the only outer $r$-range. 
In the small/medium $r$-range they are not satisfactory because the polynomial polidisperse intensities do not accurately approximate the polidisperse exact ones in the region $q<2$ (see figure 5B). 
\subsection{The 4th degree polynomial approximation}
For greater completeness we shall now briefly report the reasults that are obtained in the 
cases of  (4.1) Poisson distributions of tetrahedrons, octahedrons or cubes when the relevant 
CFs are approximated by a 4rh degree polynomial $P_4(r)$. On requires that $P_4(r)$ obeys 
constraints {\em a), b), c), d)} and {\em e)}. Then one finds that 
\begin{eqnarray}\label{6.3.1}
&&P_4(r)= 1 + r \sigma  - \frac{5 r^2 (8 \pi + 3 \pi \sigma  - 21 v)}{4 \pi} -
\frac{ r^3 (-32 \pi - 9 \pi \sigma  + 105 v)}{2 \pi}  -\nonumber \\
&&\quad\quad \quad\quad  \frac{ 7 r^4 (4 \pi + \pi \sigma  - 15 v)}{4 \pi},
\end{eqnarray}
where $v$ and $\sigma$  respectively denote the volume and the specific surface of the 
considered unit polyhedron.  The substitution of the $\sigma$ and $v$ relevant to the unit 
regular tetrahedron, octahedron and cube yields the 4th degree polynomial approximations
 of the respective CFs. The left panel of Figure 7 compares  the exact CFs to their 4th degree 
approximations, while the right panel shows the exact and the approximated form factors 
of the three particle shapes. It is evident that the agreement is far better than in Fig. 3.  
\begin{figure}[!h]
{
\includegraphics[width=7.truecm]{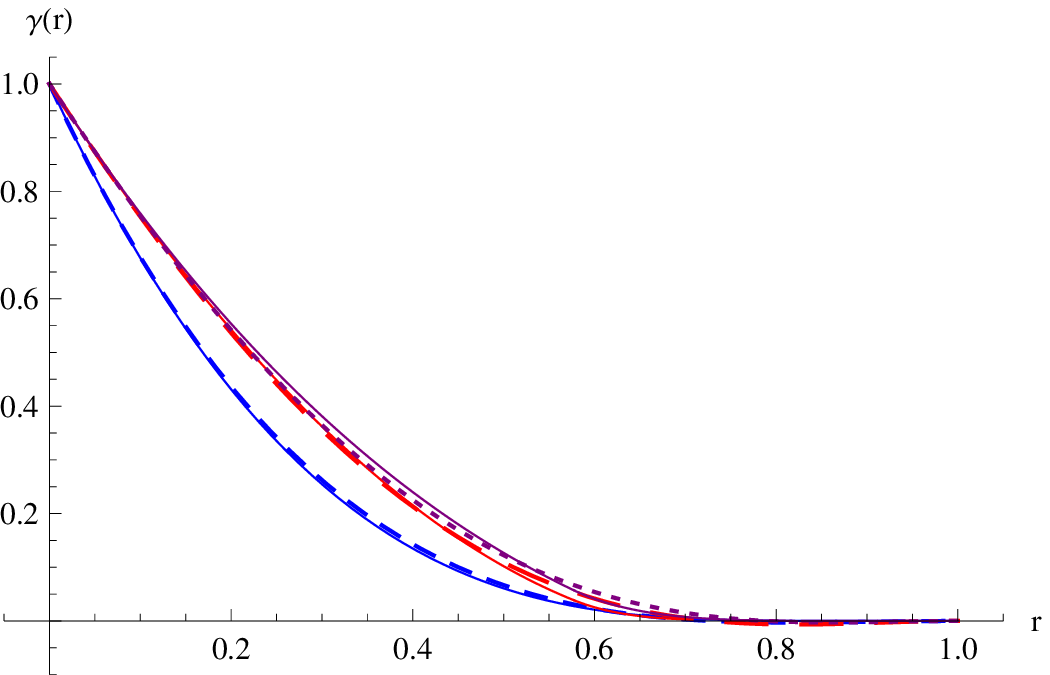}  \includegraphics[width=7.truecm]{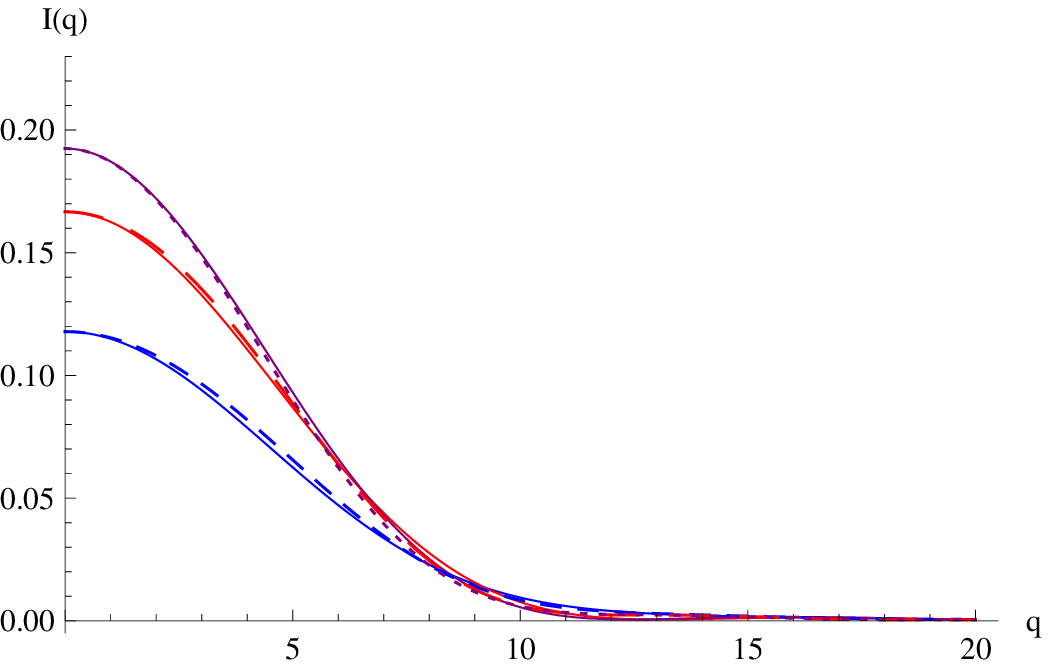} }
\caption{\label{Fig7} {Left panel: The continuous and the broken curves shows the exact and the 4th 
degree polynomially approximated  CFs of the tetrahedron (blue), octahedron (red) and cube (magenta) 
of unit maximal chord. Right panel: The FTs of the previous quantities are shown with the same symbols.}}  
\end{figure} 
The functions $G(r)$, relevant to the (4,1) Poisson polidisperse collections of tetrahedrons, octahedrons and cubes, can algebraically be evaluated by (\ref{4.2}),  (\ref{4.1c}) and (\ref{6.3.1}). They can 
also be Fourier transformed in a closed algebraic form. The results are shown in Figure 8. The  comparison with the results reported in Fig. 4 shows that the 4th degree approximation is more accurate than the 3rd degree one even though some discrepancies still survive in the range $1<q<3$ for the octahedron and cube cases.
\begin{figure}[!h]
{
\includegraphics[width=7.truecm]{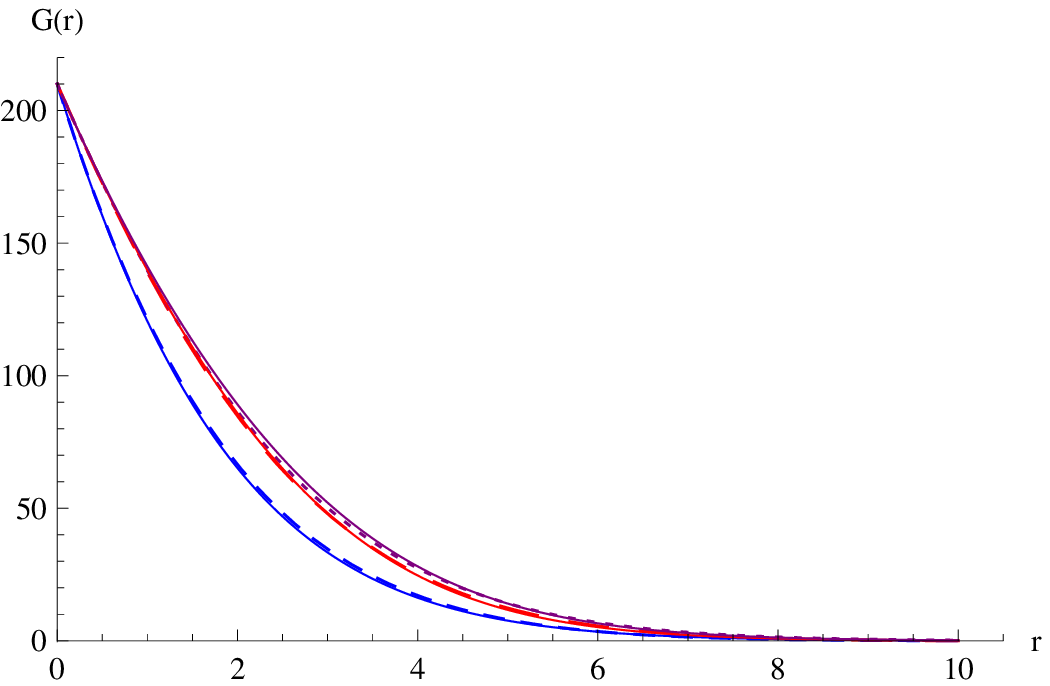} \includegraphics[width=7.truecm]{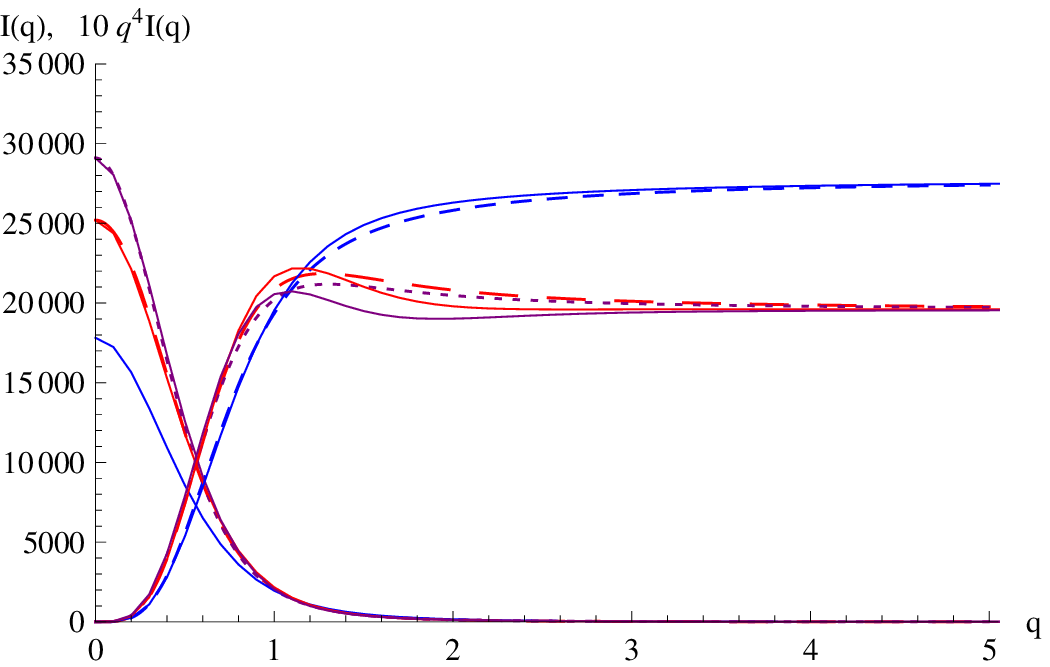} }
\caption{\label{Fig8} {Left panel: The continuous and the broken curves refer to the 
$G(r)$s obtained with the exact CFs and their 4th degree approximations. The blue, red 
and magenta  curves respectively refer to the tetrahedron, octahedron and cube  (4,1) Poisson 
collections. Right panel: behaviour of the corresponding intensities and their Porod plots.  }}
\end{figure} 
We proceed now to apply the generalized Fedorova-Schmidt method. One  has $M=4$ and
 $m=1$ so that  $\mp=2$. The resolvent equation (\ref{4.8b})  is a 2nd degree one that can 
immediately be written down because the coefficients $g_3(\sigma,v)$, $g_4(\sigma,v)$ 
and $g_5(\sigma,v)$  are obtained from (\ref{6.3.1}) according  to definition (\ref{4.3a}). 
The roots are 
\begin{eqnarray}\label{6.3.2}
\alpha_1(\sigma,v)=\frac{-\pi (64 + 15 \sigma) + 105 v +\Delta_4(\sigma,v)}{\pi (32 + 6 \sigma) - 210 v},\label{6.3.2}\\
\alpha_2(\sigma,v)=\frac{\pi (64 + 15 \sigma) - 105 v -\Delta_4(\sigma,v)}{\pi (32 + 6 \sigma) - 210 v}\label{6.3.3}
\end{eqnarray}
with 
\begeq\label{6.3.4}
\Delta_4(\sigma,v)\equiv\sqrt{\pi^2 (64 + 9 \sigma)^2 + 210 \pi (-64 + 9 \sigma) v + 11025 v^2}.
\endeq
Using the $\sigma$ and $v$ values of the considered three solids one finds 
\begeq\label{6.3.5}
\alpha_{1,t}\approx 4.55566,\quad \alpha_{1,o}\approx 1.35395,\quad \alpha_{1,c}\approx 1.47030,
\endeq 
\begeq\label{6.3.6}
\alpha_{2,t}\approx -9.32024,\quad \alpha_{2,o}\approx -8.73758,\quad \alpha_{2,c}\approx -11.9556.
\endeq
For the three cases, the resolvent kernels are immediately obtained from (\ref{5.fnlb}) recalling that $A_1(\alpha)=(\alpha_2-\alpha_1)=-A_2(\alpha)$. They have the general form ${Q_r}^3\kappa_4(Q_r)$ and $\kappa_4(Q_r)$ is a function such that its integral 
over the range $0<q<Q_M$ becomes weakly equal to zero as $Q_M\to\infty$. Then 
the integral transforms that determine the size distributions from the observed intensities  have the form of equations (\ref{6.1.8}), (\ref{5.fnlb}), (\ref{6.2.2.3}).  
Figure 9 shows the results in the cases where the scattering intensities are equal to the 
FTs of equation (\ref{4.2}) with $\gamma(r)$ equal to $P_{4,t}(r)$, $P_{4,o}(r)$ and 
$P_{4,c}(r)$ and $\cP(r)=r^3\,p(4,1,r)$. The agreement is as satisfactory as in the 
case of the 3rd degree approximation (see figure 2). 
\begin{figure}[!h]
{
\includegraphics[width=7.truecm]{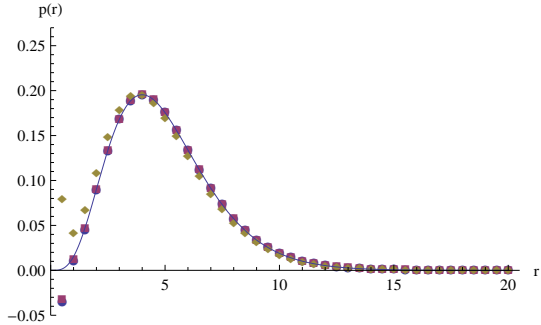} }
\caption{\label{Fig9} {The continuous curve represents the (4,1) Poisson size 
distribution and the symbols the size distributions obtained by the generalized 
Fedorova-Schmidt integral equation for the 4th degree polynomial approximations 
of the CF of the octahedron (blue full circles), the cube (red full squares) and tetrahedron (golden full diamonds). The intensities are the (4,1) polidisperse ones evaluated with the 4th degree polynomial approximations.   }}
\end{figure} \\
Figure 10 shows the resulting size distributions in the cases where the scattering 
intensities are the FTs of (\ref{4.2}) with $\gamma(r)$ equal to the exact CFs of the 
considered three platonic solids. These intensities are the observable ones. Thus 
Figure 10 shows the accuracy that can be achieved by the 4th degree approximation  
of the CFs. The comparison of Fig. 10 with  Fig. 6 shows the greater accuracy of the 
4th degree approximation. The surviving discrepancies are not large enough to make 
meaningless the application of the generalized Fedorova-Schmidt method to dilute  
polidisperse real samples of particles with the above considered polyhedral shapes. 
\begin{figure}[!h]
{
\includegraphics[width=7.truecm]{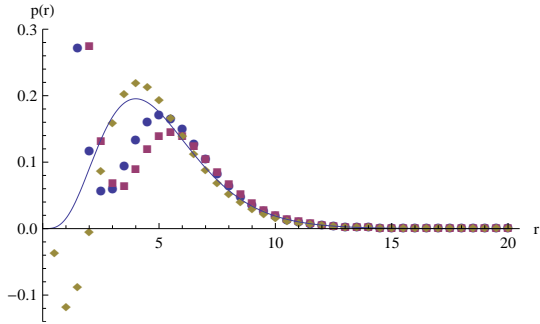} }
\caption{\label{Fig10} {Reconstruction of $p(r)$  using the 'exact'  scattering intensities  
of the  polidisperse octahedrons (blue full circles), cubes (magenta full squares) and  
tetrahedrons (golden full diamonds) and the resolvent kernels relevant to the  4th 
degree polynomial approximations. The continuos curve is the outset (4,1) Poisson 
distribution.  }}
\end{figure} 
\section{Conclusions}    
The main results of this paper are: a) a reformulation of polidisperse analysis based 
on the stick probability functions. This presentation should have made more clear the 
fact that if the system is not very dilute, the polidisperse analysis is applied to 
an intensity  that, on the average,  is somewhat smaller than the exact $I_p(q)$; 
b) the derivation of the integral transform that determines the particle size distribution 
from the scattering intensity under the assumption that  the particles have a polynomial 
CFs; c) the procedure for determining an $M$th degree  polynomial approximation to 
a given CF and finally d) the application of this procedure to  (4,1) Poisson distributions 
of  cubes,  octahedrons and tetrahedrons using the lowest significant polynomial 
approximations, i.e. the 3rd and the 4th degree ones. For the polynomial approximation 
to yield accurate results it is required that $q^4 I_{pol}(q)$ be fairly close to $q^4I_{exc}(q)$, 
where $I_{exc}(q)$ denotes the polidisperse intensity evaluated with the particle's exact 
form factor and $I_{pol}(q)$ the FT of the considered polidisperse polynomial approximation.  \\ 
  Overall the reported results indicate that polidisperse analyses of real scattering data 
might be satisfactorily  carried through with non spherical particles treated in the 
appropriate polynomial approximation. 

\appendix
\section{Derivation of equation (\ref{4.10})}
It is now proved that function $X(r)$, defined by Equation. (\ref{4.10}), is a solution of
differential equation (\ref{4.7}).\\
To this aim it is first observed that equation (\ref{4.7}) can be written as
\begin{eqnarray}
\cP^{(\mp)}(r)+A_1(r)\cP^{(\mp-1)}(r)+&\ldots&+A_{\mp}(r)\cP(r)=\label{A.1}\\
\quad && -\frac{r^{m+1}G^{(M+1)}(r)g_{m+1}}{g_{m+1}},\nonumber
\end{eqnarray}
where functions $A_1(r),\ldots,A_{\mp}(r)$ result from the explicit evaluation of the
derivatives present in (\ref{4.7}) and, after regrouping the terms, multiplying the
results by $-r^{m+1}/g_{m+1}$. Functions $Y(r,c_1,\ldots,c_{\mp})$ and
$Y\left(r,C_1(y),\ldots,C_{\mp}(y)\right)$ also are  the general integral and a particular
solution of  the homogenous differential equation associated to (\ref{A.1}) because 
 this differs  from (\ref{4.8}) by a factor. \\
The first derivative of $X(r)$ is easily obtained from Equation. (\ref{4.10}) and reads
\begin{eqnarray}
X\p(r)&=&-Y(r,C_1(r),\ldots,C_{\mp}(r))
\left(\frac{r^{m+1}G^{(M+1)}(r)}{g_{m+1}}\right)-\label{A.2}\\
\ &&\int_{r_0}^r Y\p\left(r,C_1(y),\ldots,C_{\mp}(y)\right)
\left(\frac{y^{m+1}G^{(M+1)}(y)}{g_{m+1}}\right) dy,\nonumber
\end{eqnarray}
where the prime denotes the derivative with respect to $r$. Using the
first of Equation.s (\ref{4.11}) one gets            
\begeq\label{A.3}
X\p(r)=-\int_{r_0}^r Y\p\left(r,C_1(y),\ldots,C_{\mp}(y)\right)
\left(\frac{y^{m+1}G^{(M+1)}(y)}{g_{m+1}}\right)dy.
\endeq
In a similar way, for $k=1,\ldots,(\mp-1)$, one finds that
\begeq\label{A.3a}
X^{(k)}(r)
=-\int_{r_0}^r Y^{(k)}\left(r,C_1(y),\ldots,C_{\mp}(y)\right)
\left(\frac{y^{m+1}G^{(M+1)}(y)}{g_{m+1}}\right)dy.
\endeq
The $\mp$th derivative is
\begin{eqnarray}
X^{(\mp)}(r)&=&
-\left(\frac{r^{m+1}G^{(M+1)}(r)}{g_{m+1}}\right)-\label{A.4}\\
\ &&\int_{r_0}^r Y^{(\mp)}\left(r,C_1(y),\ldots,C_{\mp}(y)\right)
\left(\frac{y^{m+1}G^{(M+1)}(y)}{g_{m+1}}\right)dy\nonumber.
\end{eqnarray}                                         
Thus one finds that
\begin{eqnarray}
\  && \sum_{k=0}^{\mp}A_k(r)X^{(k)}(r)=    
-\left(\frac{r^{m+1}G^{(M+1)}(r)}{g_{m+1}}\right)-\label{A.5}\\
\ && \int_{r_0}^r \Big[\sum_{k=0}^{\mp}A_k(r)
Y^{(k)}\left(r,C_1(y),\ldots,C_{\mp}(y)\right)\Big]
\left(\frac{y^{m+1}G^{(M+1)}(y)}{g_{m+1}}\right)dy.\nonumber
\end{eqnarray}
The expression within the square brackets is equal to zero because
$Y\left(r,C_1(y),\ldots,C_{\mp}(y)\right)$ is a solution of the homogeneous
differential equation. In this way one finds that
\begin{eqnarray}
\  && \sum_{k=0}^{\mp}A_k(r)X^{(k)}(r)=
-\left(\frac{r^{m+1}G^{(M+1)}(r)}{g_{m+1}}\right),\label{A.5}
\end{eqnarray}
and the proof that $X(r)$ is a particular integral of Equation. (A.1) is achieved.  \\
It is now proved that (\ref{4.11b}) are the solution of Equation.s (\ref{4.11}).
To this aim one puts $f_i(r)=r^{\alpha_i}$ for $i,\ldots,\mp$ and one writes
$Y(y,C_1,\ldots,C_{\mp})$ as $\sum_{i=1}^{\mp}C_i\,f_i(y)$. Equation.s (\ref{4.11})
become
\begin{eqnarray}
\sum_{i=1}^{\mp}C_i\,f_i(y)& =&0,\nonumber\\
\sum_{i=1}^{\mp}C_i\,{f_i}\p(y)& =&0\nonumber\\
\ldots &=&\ldots                  \label{A.6}\\
\sum_{i=1}^{\mp}C_i\,{f_i}^{(\mp)}(y)& =&1.\nonumber
\end{eqnarray}                                              
The $\mp\times (\mp+1)$ matrix associated to this system of linear equations is
\begin{equation*}  \left(
\begin{array}{ccccc}
f_1 & f_2 & \ldots&f_{\mp}&0 \\
{f_1}\p & {f_2}\p& \ldots&{f_{\mp}}' &0 \\
\vdots & \vdots & \ddots&\vdots&\vdots\\
{f_1}^{(\mp-1)} & {f_2}^{(\mp-1)}& \ldots&{f_{\mp}}^{(\mp-1)}&1
\end{array} \right)
\end{equation*}                      
that, using the $f_i$ expressions, becomes
\begin{equation} \label{A.7} \left(
\begin{array}{ccccc}
y^{\alpha_1} & y^{\alpha_2} & \ldots&y^{\alpha_{\mp}}&0 \\
\alpha_1\,y^{\alpha_1-1} & \alpha_2\,y^{\alpha_2-1}& \ldots&\alpha_{\mp}\,y^{\alpha_{\mp}-1}&0 \\
\vdots & \vdots & \ddots&\vdots&\vdots\\
(\alpha_1)_{(\mp-1)}\,y^{\alpha_1-\mp+1} & (\alpha_2)_{(\mp-1)}\,y^{\alpha_2-\mp+1}& \ldots&(\alpha_{\mp})_{(\mp-1)}\,y^{\alpha_{\mp}-\mp+1}&1
\end{array} \right).
\end{equation}                
The determinant $\Delta$ of the $(\mp\times\mp)$ matrix, formed by the first $\mp$  columns 
of (\ref{A.7}), is the determinant of the coefficients. It is simply evaluated observing that if one multiplies
 the terms of the $k$th row by $y^{k-1}$ for $k=1,\ldots,\mp$, the $j$th column terms have 
$y^{\alpha_j}$ as common factor. After extracting these factors, determinant $\Delta$ reduces to
to the Vendermonde determinant
\begin{equation*}
\Delta=\left(\prod_{k=1}^{\mp}y^{1-k+\alpha_k}\right){\rm det} \left|
\begin{array}{cccc}
1 & 1& \ldots&1 \\
\alpha_11 & \alpha_21& \ldots&\alpha_{\mp} \\
\vdots & \vdots & \ddots&\vdots\\
(\alpha_1)^{(\mp-1)} & (\alpha_2)^{(\mp-1)}& \ldots&(\alpha_{\mp})^{(\mp-1)}
\end{array} \right|
\end{equation*}
whose value is
\begeq\label{A.8}
\Delta=y^{ (\mp(\mp-1)/2+\sum_{i=1}^{\mp}\alpha_i) }\,\prod_{1\le\,i<j\le\mp}(\alpha_i-\alpha_j).
\endeq
To get the expression of coefficient $C_k$ one must evaluate the determinant of the $(\mp\times\mp)$ matrix obtained omitting the $k$th column  in equation (\ref{A.7}) and multiply the result by $1/\Delta$. 
The value of determinant is evaluated by the same procedure expounded above. Its value  is
\begeq
{(-1)^{k-1}y^{ ((\mp-1)(\mp-2)/2+{\sum\p}_{i=1}^{\mp}\alpha_i) }\,
{{\prod}_{1\le\,i<j\le\mp}}\p(\alpha_i-\alpha_j)}, \label{A.9}
\endeq
where the primes on the sum and product symbols denote that the sum and the product indices cannot
be equal to $k$. Dividing this result by $\Delta$ and simplifying one gets
\begeq
C_k= \frac{y^{\mp-1-\alpha_k}}{ {{\prod}\p}_j\,(\alpha_k-\alpha_j)} \label{A.10}
\endeq
that is Equation.(\ref{4.11a}).

\section{Derivation of equation. (\ref{6.1a})}   
First of all the definition of the symbols  present in Equation. (\ref{5.1a}) are as follows.
The prime on the sum denotes that index $l$ ranges over the even numbers and that
$h,\, k,\, l$ obey the constraint $h+k+l=m$;
\begeq\label{B.1}
\langle R^{2h} \rangle \equiv \frac{1}{V}\int r^{2h}\rho_p(\br)dv,\quad h=0,1,2,\ldots
\endeq
where $\rho_p(\br)$ is the characteristic function of the particle that has its gravity center
set at the origin of the Cartesian frame and its maximal chord equal to one and,finally,
\begeq\label{B.2}
a_l\equiv \frac{2^{l}}{l+1}.
\endeq
The $2m$th moment of $\gr$, using the CF definition, can be written as
\begin{eqnarray}\label{B.3}
\cG_{2m}&=&\int r^{2m}\gr\,dv =\frac{1}{V}\int r^{2m}dv\int\rho_p(\br_2)\rho_p(\br_2+\br)dv_2=\\
\quad& & \frac{1}{V}\int dv_1\int\rho_p(\br_2)\rho_p(\br_1)(\br_1-\br_2)^{2m}dv_2.\nonumber
\end{eqnarray}
One has
\begeq\label{B.4}
(\br_1-\br_2)^{2m}=(r_1^2-2\br_1\cdot\br_2+r_2^2)^m = {\sum}_{0\le h,k,l\,\le m}
\frac{m!\,(-2)^l}{h!\,k!\,l!}r_1^{2h} r_2^{2k}(\br_1\cdot\br_2)^{l}
\endeq 
with the further constraint $h+k+l=m$, 
and (\ref{B.3}) becomes
\begeq\label{B.5}
\cG_{2m}={\sum}_{0\le h,k,l\,\le m}\frac{m!\,(-2)^l}{h!\,k!\,l!V}{\cI^{2h,l}}_{a_1,\ldots,a_l}{\cI^{2k,l}}_{a_1,\ldots,a_l}
\endeq
where it has been put
\begeq\label{B.6}
{\cI^{2h,l}}_{a_1,\ldots,a_l}\equiv \int r_1^{2h}\br_{1,a_1}\ldots\br_{1,a_l}\rho_p(\br_1)dv_1
\endeq
and the convention of summing over repeated indices has been adopted. Quantity
${\cI^{2h,l}}_{a_1,\ldots,a_l}$ is a fully symmetric tensor of rank $l$ that must behave as
a scalar quantity. Hence it has the form
\begeq\label{B.7}
{\cI^{2h,l}}_{a_1,\ldots,a_l}= {\cI_0}^{2h,l}{\cS^l}_{a_1,\ldots,a_l}
\endeq
where the ${\cI_0}^{2h,l}$ expressions must be determined and ${\cS^l}_{a_1,\ldots,a_l}$ 
is a fully symmetric tensor resulting from the sum of
terms having the form $\delta_{a_1,a_{i_2}}\delta_{a_{i_3},a_{i_4}}\ldots\delta_{a_{i_{l-1}},a_{i_l}}$
where $\delta_{a,b}$ is the $(3\times 3)$ Kronecker symbol and $i_2,i_3,\ldots,i_l$ is a
permutation of $\{2,3,\ldots,l\}$ such that $i_3<i_4$, $i_5<i_6$, \ldots, and $i_{l-1}<i_l$. 
Clearly the existence of ${\cS^l}_{a_1,\ldots,a_l}$ requires that $l$ be even.
If one saturates two indices of ${\cI^{2h,l}}_{a_1,\ldots,a_l}$, from definition (\ref{B.6}) one gets
\begeq\label{B.8}
{\cI^{2h,l}}_{a,a,a_3\ldots,a_l}={\cI^{2h+2,l-2}}_{a_3\ldots,a_l}
\endeq
and by (\ref{B.7}) that
\begeq\label{B.9}
{\cI_0}^{2h,l}{\cS^l}_{a,a,a_3,\ldots,a_l}={\cI_0}^{2h+2,\,l-2}{\cS^{l-2}}_{a_3,\ldots,a_l}.
\endeq
Quantity ${\cS^l}_{a,a,a_3,\ldots,a_l}$ can be explicitly related to ${\cS^{l-2}}_{a_3,\ldots,a_l}$.
In fact from the ${\cS^l}_{\ldots}$ definition follows
\begeq\label{B.10}
{\cS^l}_{a_1,a_2,a_3,\ldots,a_l}=\delta_{a_1,a_2}{\cS^{l-2}}_{a_3,\ldots,a_l} +
\sum_{3\le k\le l}\delta_{a_1,a_k}{\cS^{l-2}}_{a_2,\ldots,a_l,{\hat k}},
\endeq
where symbol $\hat k$ means that index $a_k$ is not present. Saturating $a_1$ with $a_2$ in (\ref{B.10}),
one obtains that
\begeq\label{B.11}
{\cS^l}_{a,a,a_3,\ldots,a_l}=(3+l-2){\cS^{l-2}}_{a_3,\ldots,a_l}.
\endeq
By iteration of this relation one gets
\begeq\nonumber
{\cS^{2L}}_{a_1,a_1,a_3,a_3,a_5\ldots,a_{2L}}=
(2L+1)(2(L-1)+1){\cS^{2(L-2)}}_{a_5\ldots,a_{2L}},
\endeq
and, since ${\cS^{2}}_{a,a}=3$, one finds that
\begeq\label{B.12}
{\cS^{2L}}_{a_1,a_1,a_3,a_3,\ldots,a_{2L-1},a_{2L-1}}=\prod_{1\le j\le L}(2j+1)=\frac{2^{L+1}\Gamma(L+3/2)}{\sqrt{\pi}},
\endeq
and, by (\ref{B.6}) and  (\ref{B.7}),
\begeq\label{B.13}
{\cI^{2h,2L}}_{a_1,\ldots,a_{2L}}={\cI_0}^{2h+2L,0}{\cS^{2L}}_{a_1,\ldots,a_{2L}}\big/D(L)
\endeq
with
\begeq\label{B.14}
D(L)\equiv \frac{2^{L+1}\Gamma(L+3/2)}{\sqrt{\pi}}.
\endeq
To fully simplify the rhs of (\ref{B.5}) one must saturate ${\cI^{2h,l}}_{a,a,a_3\ldots,a_l}$
with itself. By relation (\ref{B.13}), this amounts to evaluate  ${\cS^{2L}}_{a_1,\ldots,a_{2L}}{\cS^{2L}}_{a_1,\ldots,a_{2L}}$. This quantity by (\ref{B.10})
becomes equal to (it is recalled that $l=2L$)
\begin{eqnarray}
&&3{\cS^{l-2}}_{a_3,\ldots,a_l} {\cS^{l-2}}_{a_3,\ldots,a_l}+
2\delta_{a_1,a_2}{\cS^{l-2}}_{a_3,\ldots,a_l}\sum_{3\le k\le l}\delta_{a_1,a_k}{\cS^{l-2}}_{a_2,\ldots,a_l,{\hat k}}+\nonumber\\
&&\sum_{3\le k,j\le l}\delta_{a_1,a_k}{\cS^{l-2}}_{a_2,\ldots,a_l,{\hat k}}
\delta_{a_1,a_j}{\cS^{l-2}}_{a_2,\ldots,a_l,{\hat j}}.
\end{eqnarray}
The second term is equal to $2\times2(L-1){\cS^{2(L-1)}}_{a_3,\ldots,a_{2(L-1)}}{\cS^{l-2}}_{a_3,\ldots,a_{2L}}$,
while the third term can be written as
\begin{eqnarray}
&&\sum_{3\le j\le l}\delta_{a_j,a_j}{\cS^{2(L-1)}}_{a_2,\ldots,a_{2L},{\hat j}}
{\cS^{2(L-1)}}_{a_2,\ldots,a_l,{\hat j}}+\nonumber \\
&&\sum_{3\le j\ne k\le l}\delta_{a_j,a_k}{\cS^{2(L-1)}}_{a_2,\ldots,a_{2L},{\hat k}}
{\cS^{2(L-1)}}_{a_2,\ldots,a_{2L},{\hat j}}=\nonumber\\
&&[3\times2(L-1) + (2L-3)(2L-4)]{\cS^{2(L-1)}}_{a_3,\ldots,a_{2(L-1)}}{\cS^{l-2}}_{a_3,\ldots,a_{2L}}
\nonumber
\end{eqnarray}
Collecting the above results one finds that
\begeq\nonumber
{\cS^{2L}}_{a_1,\ldots,a_{2L}}{\cS^{2L}}_{a_1,\ldots,a_{2L}}=(4L^2-1){\cS^{2(L-1)}}_{a_3,\ldots,a_{2L}}
{\cS^{2L-2}}_{a_3,\ldots,a_{2L}}.
\endeq
Iterating one gets
\begeq\label{B.15}
{\cS^{2L}}_{a_1,\ldots,a_{2L}}{\cS^{2L}}_{a_1,\ldots,a_{2L}}=\prod_{1\le j\le L}(4j^2-1)=
\frac{2^{L+\frac{1}{2}}\Gamma(L+\frac{1}{2})\Gamma(L+\frac{3}{2})}{\pi}.
\endeq
Finally, one finds that
\begeq\label{B.16}
\cG_{2m}={\sum\,'}_{0\le h,k,l\,\le m}\frac{m!(2)^{2L}}{h!\,k!\,l!\,(l+1)\,V}{\cI^{2h+2L,0}}{\cI^{2k+2L,0}}.
\endeq
Recalling definitions (\ref{B.1}) and (\ref{B.2}) the proof of (\ref{6.1a}) is completed.\\
The relation can easily be checked in the case of a spherical unit particle   
since by direct evaluation one finds, from  Equation. (\ref{6.1.1}),  that 
\begeq\label{B.17} 
\cG_{2m}= \frac{12 \pi}{72 + 108 m + 52 m^2 + 8 m^3}
\endeq 
and from (\ref{B.1}) that 
\begeq\label{B.18}
\langle R^{2h} \rangle =\frac{3\times 2^{-2 m}}{3 + 2 m}.
\endeq 
\vfill\eject
\section*{References}
\begin{description}
\item[\refup{}]Abramowitz, M. \& Stegun, I.A. (1970). {\em Handbook of Mathematical Functions}, New York: Dover. 
\item[\refup{}]Bender, C.M. \& Orszag, S. A. (1978). {\em Advanced Mathematical
Methods for Scientists and Engineers.} New York: McGraw-Hill, \S
3.3 and 3.4.
\item[\refup{}]Botet, R. \& Cabane, B.  (2012).  {\em J. Appl. Cryst.} {\bf 45},  406-416. 
\item[\refup{}]Ciccariello, S. (1984). {\em J. Appl. Phys.} {\bf 56}, 162-67.
\item[\refup{}]Ciccariello, S.  (2014). {\em J. Appl. Cryst.} {\bf 47}, in the press; arXiv: 1407.2788v1.
\item[\refup{}]Ciccariello, S.  \& Sobry,  R. (1995). {\em Acta Cryst. A} {\bf 51}, 60-69.
\item[\refup{}]Ciccariello, S., Cocco, G., Benedetti, A.  \&  Enzo, S. (19881). {\em Phys. Rev.} B{\bf 23}, 6474-6485.
\item[\refup{}] Debye, P., Anderson, H.R.  \&  Brumberger, H. (1957). {\em J. Appl.
Phys.} {\bf 20}, 679-683.
\item[\refup{}]Fedorova, I.S. \& Schmidt, P.W. (1978). {\em J. Appl. Cryst.} {\bf 11}, 405-11.
\item[\refup{}] Feigin, L.A. \&    Svergun, D.I. (1987). {\em Structure Analysis
by Small-Angle X-Ray and  Neutron Scattering}, (Plenum Press, New York).
\item[\refup{}]   Gille, W. (2013). {\em Particle and Particle Systems
characterization }, (CRC Press, London)
\item[\refup{}]Goodisman, J. (1980). {\em J. Appl. Cryst.} {\bf 13}, 132-34.
\item[\refup{}]Goodisman, J. \& Brumberger, H. (1971). {\em J. Appl. Cryst.} {\bf 4},
347-351.
\item[\refup{}]Goursat, E. (1959). {\em Differential Equations.} New York: Dover,
\S\,38 and 39.
\item[\refup{}]Guinier, A. (1946). {\em Compt. Rend.} {\bf 223}, 161-162.
\item[\refup{}]Guinier, A. \& Fournet, G. (1955). {\em Small-Angle Scattering of X-rays.} New York: John Wiley.
\item[\refup{}]Kirste, R. \& Porod, G. (1962). {\em Kolloid Z.} {\bf 184}, 1-6.
\item[\refup{}]Letcher, J.H. \& Schmidt, P.W. (1966). {\em J. Appl. Phys.} {\bf 37}, 649-655. 
\item[\refup{}]Luke, Y.L. (1969). {\em The Special Functions and Their Approximations}, Vol. I, Academic Press:New York. 
\item[\refup{}]M\'ering, J. \& Tchoubar, D. (1968). {\em J. Appl. Cryst.} {\bf 1}, 153-65.
\item[\refup{}]Moore, P.B.  (1980). {\em J. Appl. Cryst.} {\bf 13}, 168-175.
\item[\refup{}]Pedersen, J.S.  (1994). {\em J. Appl. Cryst.} {\bf 27}, 595-608.
\item[\refup{}]Porod, G. (1951). {\em Kolloid Z.} {\bf 124}, 83-114.
\item[\refup{}]Porod, G. (1967). {\em Small-Angle X-Ray Scattering. Proceedings of the 
Syracuse Conference}, edited by H. Brumberger, 1-8, Gordon \& Breach:New York.
\item[\refup{}]Roess, L.C. (1946). {\em J. Chem. Phys.} {\bf 14}, 695-697.
\item[\refup{}]Roess, L.C. \& Shull, C.G. (1947). {\em J. Appl. Phys.} {\bf 18}, 308-313.
\item[\refup{}]Taupin, D.  \& Luzzatti, V. (1982). {\em J. Appl. Cryst.} {\bf 15}, 289-300.
\end{description}
\end{document}